%% file: main.tex
\pdfoutput=1
\PassOptionsToPackage{table,x11names}{xcolor}
\documentclass[sigconf,screen,10pt]{acmart}

\usepackage{booktabs}
\usepackage{textcomp}

\usepackage[T1]{fontenc}

\input{macros}
\usepackage{lipsum}
\usepackage{tcolorbox}
\usepackage{enumitem}
\setlist{itemsep=1pt}

\newcommand{\mini}[1]{\textcolor{Blue3}{\textbf{#1}}}
\newcommand{\maxi}[1]{\textcolor{Red3}{\textbf{#1}}}

\copyrightyear{2020}
\acmYear{2020}
\setcopyright{acmlicensed}
\acmConference[SIGMOD'20]{Proceedings of the 2020 ACM SIGMOD International Conference on Management of Data}{June 14--19, 2020}{Portland, OR, USA}
\acmBooktitle{Proceedings of the 2020 ACM SIGMOD International Conference on Management of Data (SIGMOD'20), June 14--19, 2020, Portland, OR, USA}
\acmPrice{15.00}
\acmDOI{10.1145/3318464.3380564}
\acmISBN{978-1-4503-6735-6/20/06}

\begin{document}
\fancyhead{}

\title{The Solution Distribution of Influence Maximization}
\subtitle{A High-level Experimental Study on Three Algorithmic Approaches}

\author{Naoto Ohsaka}
\affiliation{\institution{NEC Corporation}}
\email{naoto.ohsaka@gmail.com}
\orcid{0000-0001-9584-4764}

\begin{abstract}
Influence maximization is among the most fundamental algorithmic problems in social influence analysis.
Over the last decade,
a great effort has been devoted to developing
efficient algorithms for influence maximization,
so that identifying the ``best'' algorithm has become a demanding task.
In SIGMOD'17, Arora, Galhotra, and Ranu
reported benchmark results on eleven existing algorithms and
demonstrated that there is no single state-of-the-art offering the best trade-off between computational efficiency and solution quality.

In this paper, we report a high-level experimental study
on three well-established algorithmic approaches for influence maximization,
referred to as Oneshot, Snapshot, and Reverse Influence Sampling (RIS).
Different from Arora et al.,
our experimental methodology is so designed that we
examine the \emph{distribution} of random solutions,
characterize the relation between the \emph{sample number} and the actual solution quality, and
avoid \emph{implementation dependencies}.
Our main findings are as follows:
\textbf{1.}~For a sufficiently large sample number,
we obtain a unique solution regardless of algorithms.
\textbf{2.}~The average solution quality of Oneshot, Snapshot, and RIS
improves at the same rate up to scaling of sample number.
\textbf{3.}~Oneshot requires more samples than Snapshot, and
Snapshot requires fewer but larger samples than RIS.
We discuss the time efficiency
when \emph{conditioning} Oneshot, Snapshot, and RIS to be of identical accuracy.
Our conclusion is that
Oneshot is suitable only if the size of available memory is limited, and
RIS is more efficient than Snapshot for large networks;
Snapshot is preferable for small, low-probability networks.

\end{abstract}

\maketitle

\input{introduction}
\input{preliminaries}

\input{review}

\input{design}

\input{results}
\input{discussion}

\input{conclusion}

\balance

\clearpage

\bibliographystyle{ACM-Reference-Format}
\bibliography{main}

\end{document}

%% file: macros.tex
\usepackage{color}
\usepackage{soul}
\usepackage{comment}
\usepackage[normalem]{ulem}

\usepackage[caption=false,font=footnotesize]{subfig}
\usepackage{multirow,multicol,dcolumn,booktabs}
\usepackage{wrapfig}

\usepackage{framed}
\usepackage{fancybox}
\definecolor{shadecolor}{gray}{0.80}

\usepackage{lscape}
\usepackage{epsfig}
\usepackage{xfrac}
\usepackage{xspace}
\usepackage{url}

\usepackage{ltxtable}

\usepackage{bm}

\newtheorem{theorem}{Theorem}[section]

\newtheorem{definition}[theorem]{Definition}

\newtheorem{problem}[theorem]{Problem}

\newcommand{\OPT}{\operatorname{OPT}}
\newcommand{\EPT}{\operatorname{EPT}}
\newcommand{\bigO}{\mathcal{O}}

\renewcommand{\Pr}{\mathbf{Pr}}
\newcommand{\E}{\mathbf{E}}

\newcommand{\inei}{\Gamma^{-}}
\newcommand{\onei}{\Gamma^{+}}
\newcommand{\ideg}{d^{-}}
\newcommand{\odeg}{d^{+}}
\newcommand{\R}{r}

\newcommand{\Inf}{\mathsf{Inf}}

\newcommand{\calG}{\mathcal{G}}

\newcommand{\calI}{\mathcal{I}}

\newcommand{\calM}{\mathcal{M}}

\newcommand{\calR}{\mathcal{R}}
\newcommand{\calS}{\mathcal{S}}

\newcommand{\bbR}{\mathbb{R}}

\newcommand{\rme}{\mathrm{e}}

\usepackage{algorithmicx}
\usepackage[ruled,section]{algorithm}
\usepackage[noend]{algpseudocode}
\algnewcommand\And{\; \textbf{and} \;}
\algnewcommand\Or{\; \textbf{or} \;}
\algnewcommand\To{\; \textbf{to} \;}
\algnewcommand\Continue{\textbf{continue}}
\algnewcommand\Not{\textbf{not}}

\algnewcommand{\algalign}[1]{\parbox[t]{\dimexpr\linewidth-\algorithmicindent}{#1\strut}}
\algrenewcommand\textproc{\textsl}

\newcommand{\karate}{\texttt{Karate}\xspace}
\newcommand{\phy}{\texttt{Physicians}\xspace}
\newcommand{\grqc}{\texttt{ca-GrQc}\xspace}

\newcommand{\BAs}{\texttt{BA\_s}\xspace}
\newcommand{\BAd}{\texttt{BA\_d}\xspace}
\newcommand{\wiki}{\texttt{Wiki-Vote}\xspace}
\newcommand{\youtube}{\texttt{com-Youtube}\xspace}
\newcommand{\pokec}{\texttt{soc-Pokec}\xspace}

\newcommand{\Barabasi}{Barab{\'a}si\xspace}

\newcommand{\oneshot}{\textsf{Oneshot}\xspace}
\newcommand{\snapshot}{\textsf{Snapshot}\xspace}
\newcommand{\ris}{\textsf{RIS}\xspace}
\newcommand{\algA}{\textsf{alg}\xspace}
\newcommand{\alg}{\textsf{alg}\xspace}

\newcommand{\UC}[1]{$\normalfont\textsc{uc}_{#1}^{}$\xspace}

\newcommand{\OWC}{{\sc owc}\xspace}
\newcommand{\IWC}{{\sc iwc}\xspace}

%% file: introduction.tex
\section{Introduction}
\label{sec:introduction}
 
Social influence among individuals plays an immense role
in decision making and information acquisition,
and the rise of online social networks has empowered it to spread out at a tremendous scale.
Understanding, predicting, and controlling social influence and its diffusion have become a big field of research called
computational social influence~\cite{chen2015computational}.
Among the most actively studied algorithmic problems in this field is
the \emph{influence maximization} problem \cite{kempe2003maximizing,kempe2005influential},
initially motivated by viral marketing~\cite{domingos2001mining}.
Conceptually, influence maximization involves identifying a small number of \emph{seed} individuals in the network
who can maximize the spread of influence.
Kempe, Kleinberg, and Tardos~\cite{kempe2003maximizing} in 2003 formulated influence maximization
as a combinatorial optimization problem on graphs, and
their framework has been broadly accepted in the research community
as well as database~\cite{cheng2013staticgreedy,cohen2014sketch,galhotra2016holistic,huang2017revisiting,nguyen2016stop,ohsaka2016dynamic,popova2018data,tang2018online,tang2015influence,tang2014influence,ke2018finding,feng2018inf2vec}.

One striking topic is the development of efficient algorithms.
Under two well-established diffusion models called
independent cascade~\cite{goldenberg2001talk} and
linear threshold~\cite{granovetter1978threshold},
Kempe et al.~\cite{kempe2003maximizing} proved that
it is NP-hard to find the optimal solution, but
the objective function referred to as the \emph{influence spread} enjoys
an excellent property called submodularity.
A natural \emph{Greedy algorithm} thus guarantees a $(1-1/\rme)$-approximation~\cite{nemhauser1978analysis}.
However, the sheer size of today's real-world networks and the stochastic nature of the diffusion process make it more challenging to execute the Greedy algorithm.
Consequently, this topic has been an active research area for the past ten-odd years
(see, e.g., \cite{chen2013information,aslay2018influence,lakshmanan2018influence}).

\subsection{Quick Review of Existing Approaches}
Our focus in this paper is on the empirical behavior of
existing algorithms for influence maximization.
Let us quickly review them (see Section~\ref{sec:review} for details).
The Greedy algorithm suffers from the intractability of evaluating the influence spread,
which is defined as the expectation over exponentially many realizations.
The requirement is thus an efficient scheme for approximating the influence spread.

The basic idea for this requirement is to construct an (unbiased) \emph{estimator} of the influence spread.
The representatives based on this idea can be classified into three approaches, namely,
\emph{\oneshot},
\emph{\snapshot}, and
\emph{Reverse Influence Sampling (\ris)}.
Each of them is parameterized by a single parameter called the \emph{sample number} that
trades between the computational complexity and the solution quality.

Up to now, there have been two different directions for realizing ``efficient'' algorithms.
On the one hand, \ris-type algorithms aim to 
determine the least sample number
required to guarantee the worst-case approximation factor theoretically.
On the other hand, \oneshot- and \snapshot-type algorithms aim to run as fast as possible for a given sample number.
Even though comparing such algorithms designed for distinct purposes is quite complicated,
\ris-type algorithms have been regarded as the state-of-the-art
supported by the near-linear time complexity ``in theory''~\cite{nguyen2016stop,huang2017revisiting,tang2014influence,tang2015influence,tang2018online}.
Note that there exist numerous heuristics that provide influence estimates quickly,
but they often result in poorly influential solutions.

\subsection{Benchmarking Study in 2017}
Arora, Galhotra, and Ranu~\cite{arora2017debunking} published a paper titled
\emph{``Debunking the Myths of Influence Maximization: An In-Depth Benchmarking Study''} at SIGMOD 2017.
The authors exposed a pitfall in comparing influence maximization algorithms by experiments
through an exhaustive benchmarking study on eleven existing algorithms
~\cite{leskovec2007cost,goyal2011celf,tang2014influence,tang2015influence,cheng2013staticgreedy,cheng2014imrank,chen2010scalablea,goyal2011simpath,jung2012irie,galhotra2016holistic,ohsaka2014fast}.
The results indicated that
\textbf{1.}~the algorithmic efficiency is sensitive to the choice of problem instances, and
\textbf{2.}~\emph{no single} state-of-the-art achieves
the best trade-off among computation time, memory consumption, and solution quality.
Unfortunately,
Arora et al.'s experimental methodology still contains several flaws
as pointed out by
Lu, Xiao, Goyal, Huang, and Lakshmanan~\cite{lu2017refutations}.
For example,
Arora et al.~\cite{arora2017debunking} used for each algorithm,
a fixed parameter value determined based on preliminary experimental results,
which does not match the research focus of \ris-type algorithms~\cite[Sect.~3.1]{lu2017refutations}.
This study complements previous studies from a different aspect.

\subsection{Our Motivations}

In this paper, we present a high-level experimental study on existing algorithmic approaches for influence maximization.
Rather than attempting to choose the best one among them,
	we would like to clarify their potential applicability.
Our key objective under this purpose is to elucidate%
\begingroup%
\addtolength\leftmargini{-2mm}%
\begin{quote}%
\emph{the empirical impact of the sample number on the solution distribution for each algorithmic approach}%
\end{quote}%
\endgroup%
\noindent as driven by the following three facts.

\begin{enumerate}[leftmargin=*,label=\textbf{\arabic*.}]
\item \textbf{Existing algorithms are randomized.}
Since an influence estimator is randomized,
each algorithm run generates random solutions as well.
Despite this nature, most of the previous studies conducted few-trials experiments only, e.g., the number of trials is
3 in \cite{tang2014influence},
5 in \cite{tang2015influence},
10 in \cite{li2017why,chen2010scalable,chen2010scalablea},
20 in \cite{huang2017revisiting},
50 in \cite{cheng2014imrank}, and
not explicitly stated in \cite{goyal2011celf,chen2009efficient,kimura2007extracting,ohsaka2014fast,cheng2013staticgreedy,cohen2014sketch,nguyen2016stop,nguyen2017social,nguyen2017billion,nguyen2017importance,galhotra2016holistic,goyal2011simpath,jung2012irie,kimura2006tractable};
conclusions based on them would be questionable.
In this paper,
we analyze the empirical distribution of random solutions made from 1,000 trials 
to gain a deeper understanding of the stochastic behavior of randomized algorithms.

\item \textbf{Sample number controls the actual solution quality.}
Selecting an appropriate sample number is crucial,
but this has been actively studied only for \ris~\cite{borgs2014maximizing,tang2014influence,tang2015influence,nguyen2017social,nguyen2017importance,nguyen2016stop,huang2017revisiting,nguyen2017billion,tang2018online}, and
thus, we have no simple way to compare
\ris-type algorithms with the other two types.
Indeed, such \emph{worst-case}  lower bounds
are too loose to explain the empirical success fully.
In this paper,
we run algorithm implementations for a wide range of sample numbers, from 1 to up to 16 million, to discover the relation between the sample number and the actual influence.
We can, for example,
find the minimum sample number required to obtain near-optimal solutions with high probability.

\item \textbf{There is a plethora of algorithm implementations.}
We aim to evaluate the algorithmic efficiency, as well;
however, a complete experimental comparison of existing implementations is hard for two reasons.
First, one can neither completely understand nor modify complex source codes published by many different research groups.
For example, on Arora et al.~\cite{arora2017debunking}'s setup,
SimPath~\cite{goyal2011simpath} got stuck in an infinite loop 
due to its different scheme for handling graph data~\cite[Sect.~3.2.2]{lu2017refutations}.
Second, we have many metrics for evaluating scalability,
e.g., CPU time and RAM usage,
which severely depend on implementations and machine configurations.
In this paper, to avoid these implementation dependencies,
we make use of simple implementations that
capture the essence of each approach.
We measure the number of vertices and edges traversed (\emph{traversal cost}) and those stored in memory (\emph{sample size}).
Remark that the former (resp.~latter) is proportional to 
running time (resp.~memory usage),
where the proportionality
constant depends on testing environments.

\end{enumerate}

\subsection{Our Findings}
Our empirical findings are summarized as follows
(see Section~\ref{sec:results} for more details).

\begin{itemize}[leftmargin=*]
    \item \textbf{Distribution of solutions}:
    We first reveal how solution distributions converge to what distribution.
    We find that
    \oneshot, \snapshot, and \ris return the unique solution for a sufficiently large sample number.
    We further find that
    \emph{
    the Shannon entropy of solution distributions
    of \oneshot, \snapshot, and \ris drops at the same rate up to scaling of sample number}.
    
    \item \textbf{Distribution of influence spread}:
    We then analyze the empirical distribution of influence spread.
    The minimum sample number required to obtain near-optimal solutions with probability 99\%
    takes a wide range of values; e.g.,
    from 64 to 8,192 for \oneshot.
    Those empirical numbers are far smaller than worst-case bounds that depend on the graph size and seed size.
    Comparing among the three approaches,
    \emph{
    the mean value improves at the same rate up to scaling of sample size}.
    To achieve the same mean influence,
    \emph{\oneshot requires up to 96 times as many samples as \snapshot requires,
    \snapshot requires ``fewer'' (say, $10^5$ times fewer) but ``larger'' (say, $10^3$ times larger) samples than \ris};
    i.e., \ris is more space-saving than \snapshot.

    \item \textbf{Computational efficiency}:
    We finally report the traversal cost.
    We find that the presence of many high-probability edges causes expensive graph traversal
    due to the emergence of a \emph{giant component}
    \cite{karp1990transitive,bollobas2001random,ohsaka2017coarsening}.
    The \emph{per-sample traversal-cost ratio among \oneshot, \snapshot, and \ris is that
    $1 : \frac{\tilde{m}}{m} : \frac{1}{n}$}, where
    $n$ is the number of vertices,
    $m$ is the number of edges, and
    $\tilde{m}$ is the sum of all edge probabilities representing the magnitude of influence,
    showing that \ris is the most per-sample time-efficient.
\end{itemize}

Section~\ref{sec:disucsion} further discusses the traversal cost
when \emph{conditioning \oneshot, \snapshot, and \ris to be of identical accuracy}.
Our conclusion is that
\textbf{1.}~\oneshot is suitable only if the size of available memory is limited, and
\textbf{2.}~\ris is more efficient than \snapshot for large complex networks;
\snapshot is preferable for small, low-probability networks.

\paragraph*{Organization}
Section~\ref{sec:pre} introduces formal definitions and known results of influence maximization.
Section~\ref{sec:review} is devoted to a systematic survey of existing algorithms.
Section~\ref{sec:design} designs our experimental methodology.
Sections~\ref{sec:results} and~\ref{sec:disucsion} report and discuss experimental results, respectively.
Section~\ref{sec:conclusion} lists future directions of this study.

%% file: preliminaries.tex
\section{Influence Maximization}
\label{sec:pre}

\subsection{Notations}
For a positive integer $\ell$, let $[\ell]$ denote the set $\{1, 2, \ldots, \ell\}$.
We deal with two types of graph.
One is a (deterministic) \emph{graph} $ G=(V,E) $, where
$V$ is a set of vertices and $E$ is a set of edges.
The other is an \emph{influence graph} $\calG = (V,E,p)$,
which captures the stochastic nature of network diffusion,
where $V$ is the vertex set, $E$ is the edge set, and
$p : E \to (0,1]$ is an \emph{influence probability} function
representing the magnitude of influence between a pair of vertices.
For a vertex $v$ in $G$,
we use $\onei_G(v)$ and $\inei_G(v)$ to denote the out-neighbors and the in-neighbors of $v$, respectively, and
we use $\odeg_G(v)$ and $\ideg_G(v)$ to denote the out-degree and the in-degree of $v$, respectively.
For a vertex set $S$, let $\R_G(S)$ denote the number of vertices reachable from $S$ in $G$.
We will omit the subscripts when $G$ is clear from the context and use the same notations for influence graph $\calG$.
We conclude this paragraph by defining influence maximization~\cite{kempe2003maximizing},
where the definition of the influence spread is deferred.

\begin{problem}[Influence maximization]
Given an influence graph $\calG=(V,E,p)$ and a seed size $k$,
the \emph{influence maximization} problem is to
find a seed set $S \subseteq V$ of size $k$
that maximizes the influence spread of $S$.
\end{problem}

\subsection{Diffusion Model}
\label{pre:model}
\emph{Network diffusion models} describe the process by which
influence (e.g., information, contamination, and virus) triggered
by a set of seed vertices spreads over the network.
In this paper, we adopt one of the well-studied models in the literature of influence maximization.
The \emph{independent cascade (IC)} model
introduced by \cite{goldenberg2001talk}
mimics the spread of infectious diseases.
In the IC model,
each vertex takes either of the two states: \emph{active} or \emph{inactive}.
An inactive vertex may become active but not vice versa.
Let us define $A_t$ as the set of newly activated vertices at discrete time step $t$, and
let $ A_{\leq t} = \bigcup_{t' \leq t} A_{t'} $.
For a \emph{seed} set $S \subseteq V$,
the vertices in $S$ are initially activated at time step $t=0$ and the others are inactive,
i.e., $A_0 = S$.
Given $A_t$ at each time step $t$, we construct $A_{t+1}$ as follows.
Each newly activated vertex $u \in A_t$ is given a single chance to
influence its inactive out-neighbors $v \not \in A_{\leq t}$, which succeeds with probability $p(u,v)$.
If this is the case,
$v$ becomes active at time step $t+1$, i.e., $v$ is added into $A_{t+1}$.
This repetition terminates within a finite time step (at most $n$), and 
$ A_{\leq n} $ is the set including all activated vertices.

The \emph{influence spread} of a seed set $S$ in $\calG$,
denoted $\Inf_{\calG}(S)$,
is defined as the expected number of activated vertices by
initially activating seed vertices in $S$,
i.e., 
$
\Inf_{\calG}(S) = \E[|A_{\leq n}|].
$
Since $\Inf_{\calG}(\cdot)$ can be viewed as a function
on a subset of vertices, we call
$\Inf_{\calG}: 2^V \to \bbR_{\geq 0}$ an \emph{influence function}.

Here, we describe the \emph{random-graph interpretation}~\cite{kempe2003maximizing} 
that characterizes the IC model.
For an influence graph $\calG = (V,E,p)$,
consider the distribution over deterministic graphs $(V,E')$,
where $E'$ is obtained from $E$ by maintaining each edge $e$ with probability $p(e)$.
We use $ G \sim \calG $ to mean that $G$ is a \emph{random graph} sampled from this distribution.
Then, the influence spread of a seed set $S$ in $\calG$ is equal to
the expected number of vertices reachable from $S$
in $G \sim \calG$, i.e., $\Inf_{\calG}(S) = \E_{G \sim \calG}[\R_{G}(S)]$.
This fact tells us that
we do not have to consider the chronological order of activation trials,
but we need to consider reachability on random graphs.

\input{tab/review-taxonomy}

\subsection{Intractability and Approximability}
Let us recall the intractability.
Formally, it has been proven that
for the IC model, influence maximization is NP-hard to solve exactly~\cite[Theorem 2.4]{kempe2003maximizing}, and
it is even $\sharp$P-hard to compute the influence spread exactly~\cite[Theorem 1]{chen2010scalable}.

Albeit the negative results, we can obtain approximation.
The striking result of Kempe et al.~\cite[Theorem 2.2]{kempe2003maximizing} is that
the influence function is monotone and submodular.
Here, a set function $ f:2^V \to \bbR $ is said to be
\emph{monotone} if it holds that $ f(S) \leq f(T) $ whenever $ S \subseteq T \subseteq V $ and
\emph{submodular} if it holds that $ f(S+v) - f(S) \geq f(T+v)-f(T) $ whenever $ S \subseteq T \subseteq V, v \in V \setminus T $.
The classical result on submodular functions
by Nemhauser, Wolsey, and Fisher~\cite{nemhauser1978analysis} then tells us that
the simple Greedy, which iteratively adds an element that makes the maximum marginal increase in function value, produces a $(1-1/\rme)$-approximate solution, i.e.,
it holds that $f(S) \geq (1-1/\rme) f(S^*)$, where
$S$ is the $k$-seed Greedy solution and
$S^*$ is the optimal solution of size $k$.

Putting it all together, we have that Greedy achieves a constant-factor approximation in polynomial time.
The influence spread can be approximated within a factor of $(1 \pm \epsilon)$ for any $\epsilon > 0$ by running  simulations $\Omega(\epsilon^{-2}n^2)$ times~\cite{kempe2015maximizing}.

%% file: tab/review-taxonomy.tex
\begin{table*}[htbp]
\caption{Three popular algorithmic approaches for influence maximization.}
\label{tab:review:taxonomy}
\rowcolors{2}{gray!20}{white}
\centering
\small
{\fontsize{8}{8}\selectfont
\begin{tabular}{c|cccccc} \toprule
\textbf{approach} &
\textbf{sample number} &
\textbf{sample size} &
\multicolumn{2}{c}{\textbf{exp.~traversal cost} (at $k=1$)} &
\textbf{time complexity}
 \\
&
(typical value) &
$=$ (\# vertices) $+$ (\# edges) &
vertex &
edge &
(naive impl.)
\\
\midrule

\oneshot & \# simulations $\beta$ ($10^4$) &
    -- &
	$ \beta \sum_{v} \Inf_{\calG}(v) $ &
	$ \leq \beta m \max_{v} \Inf_{\calG^\top}(v) $ &
    $\bigO(\beta knm)$
    \\
\snapshot & \# random graphs $\tau$ ($10^2$) &
    $\tau \tilde{m}$ in exp. ($ \tilde{m} := \sum_{e} p(e) $) &
    $ \tau \sum_{v} \Inf_{\calG}(v) $ &
    $ \leq \tau m \max_{v} \Inf_{\calG^\top}(v) $ &
    $\bigO(\tau knm)$
	\\
\ris & \# RR sets $\theta$ (n/a) &
    $ \theta \EPT $ in exp. ($\EPT := \frac{1}{n} \sum_{v} \Inf_{\calG}(v)$) &
	$ \theta \frac{1}{n} \sum_{v} \Inf_{\calG}(v) $ &
	$ \leq \theta \frac{m}{n} \max_{v} \Inf_{\calG}(v) $ &
	$\bigO(\theta \frac{m}{n} \max_{v} \Inf_{\calG}(v))$
    \\
\bottomrule
\end{tabular}
}
\end{table*}

%% file: review.tex
\input{tab/review-rep}

\section{Algorithms Review}
\label{sec:review}

In this section, we review existing algorithms for influence maximization systematically.
Based on the mechanism of influence estimation,
we partition existing algorithms into three approaches
shown in Table~\ref{tab:review:taxonomy}, namely,
\emph{\oneshot} (Section~\ref{sec:review:one-shot}),
\emph{\snapshot} (Section~\ref{sec:review:snapshot}),
\emph{Reverse Influence Sampling (\ris)} (Section~\ref{sec:review:ris}),
and others (Section~\ref{sec:review:others}).

\subsection{Kempe et al.'s Greedy Algorithm}
\label{sec:review:greedy}

Before taking up each approach,
we explain the Greedy algorithm of \cite{kempe2003maximizing}.
It begins with an empty solution and iteratively adds an element that
makes the largest marginal increase in influence into the solution until $k$ elements have been added.
Remark that the influence function is evaluated for at most $ nk $ seed sets.
Due to monotonicity and submodularity of the influence function,
the resulting solution is a $(1-1/\rme)$-approximation~\cite{nemhauser1978analysis};
this factor is the best possible~\cite{feige1998threshold}.
By contrast, it often provides near-optimal (e.g., 0.95 times the optimum) solutions empirically~\cite{krause2008near,sharma2015greedy,leskovec2007cost}.

The limitation of Kempe et al.'s Greedy algorithm is that
we are unable to evaluate $\Inf(\cdot)$ exactly.
Fortunately,
even if we are given only an approximate value oracle,
which approximates the actual value
within a factor of $(1 \pm \epsilon$),
running Greedy on this achieves a $(1-1/\rme-\bigO(k\epsilon))$-approximation~\cite{horel2016maximization}.
Monte-Carlo simulations can be employed for this purpose; however, running them for $nk$ seed sets is computationally prohibitive,
e.g., it takes a few days on graphs with ten thousand vertices~\cite{kimura2006tractable}.
Subsequent studies thus seek efficient and accurate methods for estimating the influence spread.

\input{tab/review-framework}

\subsection{A Simple Greedy Framework}

Here, we introduce a simple greedy framework to describe existing algorithms in a unified manner.
Most of the existing algorithms fall into this framework, and
Table~\ref{tab:review:rep} lists up representative algorithms
that belong to either of the three approaches.
Our greedy framework shown in Algorithm~\ref{alg:review:framework}
requires the following three procedures.
Hereafter, we denote the seed vertex chosen at the $\ell$-th iteration by $v_\ell^{}$.

\begin{itemize}[leftmargin=*]
\item \textproc{Build}($\calG, \textit{sample number}$)
	builds an estimator for the influence function given
	influence graph $\calG$ with an approach-specific parameter,
	called ``\emph{sample number}'' in this paper.
\item \textproc{Estimate}($S_{\ell-1}, v$)
	returns an estimate for the marginal influence of $v$ with respect to $S_{\ell-1}$
	(i.e., $ \Inf(S_{\ell-1}+v) - \Inf(S_{\ell-1}) $), or
    the influence spread of $S_{\ell-1} + v$.
    The results will be the same regardless.
\item \textproc{Update}($v_{\ell}^{}$)
	updates the current estimator given the next seed $v_{\ell}^{}$ to be able to
    estimate the marginal influence with respect to the latest seed set
    $ S_{\ell} := S_{\ell-1} + v_{\ell}^{} $.
\end{itemize}

In what follows,
we clarify the concept and features of each algorithmic approach and
explain how to implement \textproc{Build}, \textproc{Estimate}, and \textproc{Update}  in a simple manner.
We then analyze the computational efficiency.
Besides the asymptotic complexity,
we measure the computational effort in terms of how many times vertices and edges were touched.
Concretely speaking, the \emph{vertex (resp.~edge) traversal cost} is defined as
the number of vertices (resp.~edges) examined (possibly more than once), and
the \emph{sample size} is defined as
the number of vertices and edges stored as
approach-specific samples in memory.
Traversal cost and sample size are implementation independent and approximately proportional to
the running time and memory usage, respectively.
Further, we present a brief survey on existing ``efficient'' implementations.

There are two types of ``efficient.''
One means that the sample number is \emph{as small as possible} while satisfying a specified precision requirement,
which demands a good criterion for sample number determination.
The other means that for a given sample number (which may be determined based on the above-mentioned criterion),
the implementation runs \emph{as fast as possible},
which demands the complexity reduction.
We note that
\oneshot- and \snapshot-type algorithms have focused on
the latter kind of efficiency while
\ris-type algorithms have focused on the former kind of efficiency
as will be clarified.
This discrepancy would be
the reason why previous experimental comparisons led to questionable claims.

\input{tab/review-one-shot}

\subsection{Oneshot Algorithms}
\label{sec:review:one-shot}

\subsubsection{Concept}
\emph{\oneshot-type algorithms}
(a.k.a.~simulation-based~\cite{li2018influence}) execute Monte-Carlo simulations of the diffusion process many times on the spot
when estimates are needed.
Algorithm~\ref{alg:review:one-shot} shows pseudocode of \oneshot, of which sample number $\beta$ specifies the \emph{number of simulations} to be performed.
In \textproc{Estimate}, given a seed set, we simulate the diffusion process $\beta$ times and
return the average number of activated vertices.
This estimate is unbiased.
Nothing is done in \textproc{Build} and \textproc{Update}.
One concern is that neither submodularity nor monotonicity is guaranteed~\cite{cheng2013staticgreedy}
because \textproc{Estimate}'s return values are independent of each other.

\subsubsection{Computational Complexity}
We analyze the computational complexity of \oneshot.
Each call of \textproc{Estimate} for a seed set $S$
examines $\Inf(S)$ vertices in expectation and their outgoing edges.
In particular, the vertex and edge traversal cost at $k=1$ is
$\beta \sum_v \Inf(v)$ and $\beta m \max_v \Inf_{\calG^\top}(v)$, respectively,
whose proof is deferred to Appendix.
Note that the entire computational complexity is  bounded by $ \bigO(\beta kmn) $.
In \textproc{Estimate}, we store $|A_{\leq n}| \leq n$ vertices, which is negligible.

\subsubsection{Efficient Implementations}
\paragraph{\textproc{Estimate} Call Pruning}
We then review existing implementation techniques.
Since simulating network diffusion is computationally expensive,
reduction of \textproc{Estimate} calls is a promising direction.
A convenient technique uses \emph{upper bounds} for the marginal influence to
identify vertices that would never be selected as a seed.
Such upper bounds are quickly derived without simulations
by utilizing submodularity \cite{leskovec2007cost,goyal2011celf} and linear systems \cite{zhou2013ublf,zhou2014upper}.

\paragraph*{Sample Number Determination}
Another direction is to determine an appropriate sample number $\beta$
given a precision requirement.
Theoretically,
setting $ \beta = \Omega( \epsilon^{-2} k^2 n (\log \delta^{-1} + \log k) / \OPT_k ) $ for any $\epsilon > 0$,
where $ \OPT_k = \max_{S: |S|=k} \Inf(S) $,
achieves
a $(1-1/\rme-\epsilon)$-approximation with probability $1-\delta$~\cite[Lemma 10]{tang2014influence};
however, none of the existing algorithms adopt such bounds
due to the inefficiency of estimating $ \OPT_k $.
In contrast, tens of thousands of simulations are sufficient to obtain reasonable solutions in practice~\cite{kempe2003maximizing,cheng2013staticgreedy}.

\input{tab/review-snapshot}

\subsection{Snapshot Algorithms}
\label{sec:review:snapshot}

\subsubsection{Concept}
\emph{\snapshot-type algorithms}
generate random graphs from an influence graph in advance and
	share them over the entire greedy seed selection, which
	makes room for a bunch of speed-up techniques.
Algorithm~\ref{alg:review:snapshot} shows pseudocode of \snapshot algorithms,
of which sample number $\tau$ specifies the \emph{number of random graphs} to be generated.
In \textproc{Build}, we independently sample
$\tau$ random graphs, denoted $G^{(1)}, G^{(2)}, \ldots, G^{(\tau)}$, from $\calG$.
An unbiased estimator of the influence spread of seed set $S \subseteq V$ is then defined as
$
	\frac{1}{\tau} \sum_{i \in [\tau]} \R_{G^{(i)}}^{}(S).
$
Hence, in \textproc{Estimate}, we take the size of reachable sets and return the average.
\textproc{Update} does nothing by default.
Unlike \oneshot, this estimator enjoys both monotonicity and submodularity since $G^{(i)}$'s are fixed~\cite{cheng2013staticgreedy}.

\subsubsection{Computational Complexity}
\textproc{Build} touches each edge only $\tau$ times, which does not dominate the whole time complexity.
In \textproc{Estimate}, for each random graph,
we scan vertices reachable from the seed set and their outgoing edges
(e.g., by a breadth-first search).
Hence, the expected vertex (resp.~edge) traversal cost at $k=1$ is up to
$\tau \sum_v \Inf(v)$ (resp.~$\tau m \max_v \Inf_{\calG^\top}(v) $),
whose proof is similar to that of \oneshot.
Remark that the entire running time is bounded by $\bigO(\tau kmn)$.
The sample size is expected to be $\tau \tilde{m}$,
where $ \tilde{m} = \sum_e p(e) $
is the expected number of edges in $G \sim \calG$.

\subsubsection{Efficient Implementations}
\paragraph{Fast Reachability Computation}
Besides applying the reduction technique for \oneshot,
\emph{quick reachability computation} in \textproc{Estimate} seems helpful.
Here, we explain a graph reduction technique in
\cite{kimura2007extracting,kimura2010extracting,ohsaka2014fast}.
Consider the situation where we are about to choose the second seed.
In \textproc{Update}($v_1^{}$),
we construct the subgraph $ H^{(i)} $ obtained from  $G^{(i)}$
by deleting vertices reachable from the first seed $v_1^{}$ and their incident edges.
Then,
it follows that 
$ \R_{G^{(i)}}(\{v_1^{}, v\}) - \R_{G^{(i)}}(v_1^{}) = \R_{H^{(i)}}(v) $
for any vertex $v$.
Therefore, it suffices to traverse on smaller $ H^{(i)} $.
Repeatedly applying this reduces the traversal cost in the subsequent iterations without disturbing estimates.

Meanwhile, finding the most influential seed (i.e., the iteration at $\ell=1$) is still laborious.
In the first iteration, we need to compute the number of vertices reachable from every vertex.
This computation problem is precisely the \emph{descendant counting problem}~\cite{cohen1997size} and
unsolvable in truly-subquadratic time, assuming the strong exponential time hypothesis~\cite{borassi2016note}.
Previous work used fast approximation or heuristics, e.g.,
reachability sketches~\cite{cohen1997size,chen2009efficient},
bottom-$k$ min-hash sketches~\cite{cohen2014sketch}, and
pruned breadth-first searches~\cite{ohsaka2014fast}.

\paragraph*{Sample Number Reduction}
Since the sample number $\tau$ governs not only the computation time but also the memory consumption unlike \oneshot,
it is more crucial to determine the value of $\tau$ appropriately.
Generating
$ \tau = \Omega(\frac{n^2}{\epsilon^2} (k \log n + \log \delta^{-1})) $
random graphs suffices to ensure that the seed set obtained
has influence at least $ (1-1/\rme)\OPT_k - \epsilon $ with probability at least $1-\delta$~\cite[Prop.~3]{karimi2017stochastic}.
Perhaps surprisingly,
it was empirically observed 
\cite{kimura2007extracting,kimura2010extracting,cheng2013staticgreedy} that
\snapshot requires fewer random graphs (e.g., $\tau=100$) than
the number of simulations required by \oneshot (e.g., $\beta=10{,}000$),
which might be unpredictable from their principles.
One reason is that
\snapshot's estimator preserves monotonicity and submodularity
while \oneshot's estimator does not so.
However, to the best of our knowledge, no clear theoretical justification for such empirical reduction in sample number has been made.

\input{tab/review-ris}

\subsection{Reverse Influence Sampling}
\label{sec:review:ris}

\subsubsection{Concept}
\emph{Reverse Influence Sampling (\ris)} (a.k.a. sketch-based \cite{li2018influence}) pioneered by Borgs, Brautbar, Chayes, and Lucier \cite{borgs2014maximizing},
is the first near-linear time algorithm (for fixed $k$) for influence maximization.
The elegant insight is
the reduction of influence maximization to stochastic maximum coverage,
which is efficiently approximable.
The critical notion of serving as a bridge between them is \emph{reverse reachable sets}.
\begin{definition}[Reverse reachable set \cite{borgs2014maximizing}]
	For a target $z \in V$, a \emph{reverse reachable (RR) set} for $z$ under IC
	is defined as a set of vertices that can reach $z$ in random $G \sim \calG$.
	An RR set for a random target, denoted $R$, is referred to as an \emph{RR set}.
\end{definition}
Then, we have that $ \Pr_R[R \cap S \neq \emptyset] = \frac{\Inf(S)}{n} $ for any vertex set $S \subseteq V$~\cite[Observation 3.2]{borgs2014maximizing},
which intuitively means that influential vertices frequently appear in RR sets.

Influence maximization is therefore equivalent to
	a maximum coverage problem on exponentially many RR sets.
To solve it approximately,
\ris polls a few RR sets and
runs Greedy on them to return an approximate solution.
Let $\calR$ be a collection of RR sets, and
	we define $ F_\calR(S) $ as the fraction of RR sets in $\calR$ intersecting $S$, i.e.,
$
	F_{\calR}(S) = \frac{|\{ R \in \calR \mid R \cap S \neq \emptyset \}|}{|\calR|}.
$
We then have $n \cdot F_{\calR}(S)$ an unbiased estimate for $\Inf(S)$.

Algorithm~\ref{alg:review:ris} shows pseudocode of \ris,
	which is characterized by the number $\theta$ of RR sets to be drawn.
\textproc{Build} constructs a collection $\calR$ of $\theta$ RR sets.
For IC, an RR set can be found efficiently by a reverse simulation~\cite{borgs2014maximizing,tang2015influence}.
\textproc{Estimate} computes the marginal coverage of vertex $v$ as $F_{\calR}(v)$.
\textproc{Update} given the new seed $v_{\ell}$
removes RR sets including $v_{\ell}$ from $\calR$ so that
$ F_{\calR}(v) $ is the marginal coverage with respect to $S_{\ell}$.

\subsubsection{Computational Complexity}
Remark that the number of steps taken by generating an RR set is
the sum of in-degrees of vertices in the set~\cite{borgs2014maximizing}.
The \emph{weight} of an RR set $R$, denoted $w(R)$, is defined as
$ \sum_{v \in R} \ideg(v) $.
By definition,
an RR set contains vertex $v$ with probability $\Inf(v) / n$, and
it turns out that
the expected size of an RR set is $ \EPT := \E[|R|] = \sum_{v} \Inf(v) / n $ (corresp.~to the vertex traversal cost).
The expected number of steps taken by generating an RR set is
$ \E[w(R)] $, which is bounded from above by $\frac{m}{n} \max_{v}\Inf(v)$ (corresp.~to the edge traversal cost).
Accordingly,
\textproc{Build} finishes in $ \bigO(\theta \frac{m}{n} \max_{v}\Inf(v)) $ expected time, and so does the entire algorithm,\footnote{See \cite[Theorem 3.1]{borgs2014maximizing} for fast implementations of \textproc{Estimate} and \textproc{Update}.} and
the sample size is bounded by $\theta \EPT$.
Here, it holds that
$ \EPT \leq 1 + \tilde{m}, $
i.e.,
the sample size of \ris is less than that of \snapshot
if $\theta=\tau$ up to addition by one,
of which proof is deferred to Appendix.

\subsubsection{Efficient Implementations}
\label{sec:review:ris:impl}

\paragraph*{Sample Number Determination}
Unlike the case of \oneshot and \snapshot,
most of the research on \ris focus on a proper selection of sample number $\theta$, or equivalently,
a stopping condition for RR-set generation.
The standard requirement is to draw as few RR sets as possible
that yield a ``theoretical worst-case guarantee'' on
a $(1-1/\rme-\epsilon$)-approximation
with probability $1-\delta$ 
\cite{borgs2014maximizing,tang2014influence,tang2015influence,nguyen2017social,nguyen2017importance,nguyen2016stop,huang2017revisiting,nguyen2017billion,tang2018online}.

Borgs et al.~\cite{borgs2014maximizing,borgs2016maximizing} first showed that
terminating RR-set generation when the total weight exceeds $\bigO(\epsilon^{-2} k (m+n) \log n)$,
which implies that $ \theta \approx \bigO( \frac{\epsilon^{-2} k n \log n}{\OPT_k} ) $
(this factor is $k$ times smaller than \oneshot's upper bound~\cite{tang2014influence}),
we have the desired guarantee.
The whole process thus can be done in almost linear time,
which is runtime-optimal up to a logarithmic factor.
However, a hidden constant sorely limits the usability against large graphs, and
subsequent research has aimed at designing tight stopping criteria
such as based on
optimal influence bounds~\cite{tang2014influence,tang2015influence},
degree-based thresholding~\cite{nguyen2017social,nguyen2017billion}, and
search-and-verify~\cite{nguyen2016stop,li2017why,huang2017revisiting}.
It is worth mentioning that these algorithms ``have roughly the same running time when their \uwave{empirical accuracies} are the
same''~ \cite[Sect.~3.1]{lu2017refutations}
(wavy underline was added by the author).
This is not the case for \oneshot and \snapshot.

\paragraph*{Complexity Reduction}
A few reduction techniques exist,
e.g., graph compression~\cite{popova2018data} and importance sampling~\cite{nguyen2017importance}.

\subsection{Other Approaches}
\label{sec:review:others}

We review two approaches not tested in this paper.

\paragraph*{Binary Decision Diagrams for Exact Computation}
Exact computation of the influence spread has been studied though it consumes exponential time.
The current fastest exact algorithm is the one designed by
Maehara, Suzuki, and Ishihata~\cite{maehara2017exact}.
The idea is to make use of the \emph{binary decision diagram}~\cite{bryant1986graph}
for computing the two-terminal network reliability, which is $\sharp$P-hard~\cite{valiant1979complexity}.
Maehara et al.'s algorithm was successful with
computing the exact influence on graphs with up to a hundred edges.

\paragraph*{Heuristics for Quick Guesses}
To avoid expensive sampling procedures,
there have been developed cheap heuristics at the price of estimation accuracy.
One popular approach is to make a strong assumption on the process of network diffusion under which the influence spread can be computed quickly, e.g.,
influence follows along with shortest paths~\cite{kimura2006tractable} or trees~\cite{chen2010scalable}, and
the zone of influence is bounded by neighbors~\cite{chen2009efficient}.
Others employ
linear systems~\cite{jung2012irie,cheng2014imrank} and
graph reduction~\cite{purohit2014fast,mathioudakis2011sparsification,ohsaka2017coarsening}.
Such heuristics are faster than the three approaches, but
resulting seed sets have less influence.

%% file: tab/review-rep.tex
\begin{table}[tbp]
\caption{Representative existing algorithms.}
\label{tab:review:rep}
\rowcolors{2}{gray!20}{white}
\centering
{\fontsize{7.5}{7.5}\selectfont
\begin{tabularx}{\linewidth}{c|X} \toprule
\textbf{approach} &
\textbf{representatives} \\ \midrule
\oneshot & 
	CELF~\cite{leskovec2007cost},
	CELF++~\cite{goyal2011celf},
	UBLF~\cite{zhou2013ublf,zhou2014upper},
	SIEA~\cite{nguyen2017outward,nguyen2017outwarda}.
    \\
\snapshot &
    Bond Percolation~\cite{kimura2007extracting,kimura2010extracting},
	NewGreedy~\cite{chen2009efficient},
	MixedGreedy~\cite{chen2009efficient},
	StaticGreedy~\cite{cheng2013staticgreedy},
	PMC~\cite{ohsaka2014fast},
	SKIM~\cite{cohen2014sketch}.
	\\
\ris &
    RIS~\cite{borgs2014maximizing}
	TIM$^+$~\cite{tang2014influence},
	IMM~\cite{tang2015influence},
	LISA~\cite{dinh2015social,nguyen2017social},
	BCT~\cite{nguyen2016cost,nguyen2017billion},
	SSA~\cite{nguyen2016stop},
	SSA-Fix~\cite{huang2017revisiting},
    WebGraph framework~\cite{popova2018data},
    OPIM~\cite{tang2018online}.
    \\ \bottomrule
\end{tabularx}
}
\end{table}

%% file: tab/review-framework.tex
\begin{algorithm}[tbp]
\caption{Simple greedy framework.}
\label{alg:review:framework}
{\fontsize{7.5}{7.5}\selectfont
\begin{algorithmic}[1]
\State call \Call{Build}{$ \calG = (V,E,p)$, \textit{sample number}} \textbf{and} $S_0 \leftarrow \emptyset$.
\State randomly shuffle the order of vertices in $V$.
\ForAll{$\ell = 1, 2, \ldots, k$}
    \State call \Call{Estimate}{$S_{\ell-1}, v$} \textbf{for all} $v \in V$.
	\State $ v_\ell^{} \leftarrow $ last vertex with maximum (marginal) influence.
    \State call \Call{Update}{$ v_\ell^{} $} \textbf{and} $ S_{\ell} \leftarrow S_{\ell-1} + v_\ell^{} $.
\EndFor
\State \Return $S_k$.
\end{algorithmic}
}
\end{algorithm}

%% file: tab/review-one-shot.tex
\begin{algorithm}[tbp]
\caption{Naive implementation of \oneshot.}
\label{alg:review:one-shot}
{\fontsize{7.5}{7.5}\selectfont
\begin{algorithmic}[1]
\Procedure{Build}{$\calG = (V,E,p), \beta$} \Comment{do nothing.}
\EndProcedure
\Procedure{Estimate}{$S_{\ell-1}, v$}
    \For{$\beta$ times}
    	\State simulate the diffusion model from $S_{\ell-1} + v$ in Sect.~\ref{sec:review:one-shot}.
    	\State count \# activated vertices.
    \EndFor
    \State \Return total \# activated vertices divided by $\beta$.
\EndProcedure
\Procedure{Update}{$v_{\ell}^{}$} \Comment{do nothing.}
\EndProcedure
\end{algorithmic}
}
\end{algorithm}

%% file: tab/review-snapshot.tex
\begin{algorithm}[tbp]
\caption{Naive implementation of \snapshot.}
\label{alg:review:snapshot}
{\fontsize{7.5}{7.5}\selectfont
\begin{algorithmic}[1]
\Procedure{Build}{$\calG = (V,E,p), \tau$}
	\State generate $\tau$ random graphs $G^{(1)}, \ldots, G^{(\tau)}$ from $\calG$.
\EndProcedure
\Procedure{Estimate}{$S_{\ell-1}, v$}
	\State \Return $ \frac{1}{\tau} \sum_{i \in [\tau]} \R_{G^{(i)}}(S_{\ell-1}+v) - \R_{G^{(i)}}(S_{\ell-1}) $.
\EndProcedure
\Procedure{Update}{$v_\ell^{}$} \Comment{do nothing.}
\EndProcedure
\end{algorithmic}
}
\end{algorithm}

%% file: tab/review-ris.tex
\begin{algorithm}[tbp]
\caption{Naive implementation of \ris.}
\label{alg:review:ris}
{\fontsize{7.5}{7.5}\selectfont
\begin{algorithmic}[1]
	\Procedure{Build}{$\calG=(V ,E, p), \theta$}
		\State $ \calR \leftarrow \emptyset $.
		\For{ $\theta$ times }
			\State $ R \leftarrow $ an RR set under $ \calM $ \textbf{and} $ \calR \leftarrow \calR + R $.
		\EndFor
	\EndProcedure
    \Procedure{Estimate}{$S_{\ell-1}, v$}
    	\State \Return $ n \cdot F_{\calR}(v). $
	\EndProcedure
    \Procedure{Update}{$v_\ell^{}$}
		\State remove RR sets including $v_\ell^{}$ from $\calR$.
    \EndProcedure
\end{algorithmic}
}
\end{algorithm}

%% file: design.tex
\section{Design}
\label{sec:design}

In this section, we design our experimental methodology.
We recapitulate two features of the algorithms reviewed so far.
First,
\oneshot, \snapshot, and \ris run Greedy on the respective influence estimator to yield random solutions.
The issue is that
examining the results of a few trials may not tell much about mean, tail, and diversity accurately.
Our strategy is thus to construct the \emph{distribution} of seed sets and discover its structural properties.
Second, whereas the sample number governs the influence spread,
we have no way of choosing a sample number of \oneshot and \snapshot according to a given precision requirement, and
worst-case approximation factors are less than $0.64$, which is too loose to explain the empirical success adequately.
Our strategy is thus to discover the relation between
the sample size and the \emph{actual} influence spread
rather than worst-case approximation factors.

To realize these,
we run algorithm \algA (e.g., \oneshot) with sample size $s$ (e.g., $\beta$) $T$ times, and
we then record each obtained seed set and its influence spread
to empirically construct
the seed set distribution  $\calS^{(s)}$ and the influence distribution $\calI^{(s)}$.
We can, for example, understand the limit behavior by
verifying the convergence of $\calS^{(s)}$ for a large $s$,
understand the tail behavior by
analyzing the first percentile of $\calI^{(s)}$, and
compare \oneshot, \snapshot, and \ris,
assuming that the sample size was ideally chosen.

In the following,
we contemplate the selection of
\textbf{1.}~algorithm implementations,
\textbf{2.}~network data, and
\textbf{3.}~edge probabilities
to grasp the behavioral trends of each approach.

\subsection{Algorithm Implementations}
Since \oneshot, \snapshot, and \ris use random numbers,
we here clarify where to invoke what type of
pseudorandom number generator (PRNG).
For each algorithm run,
we use different seed values to initialize the state of a PRNG to obtain randomized solutions.
The implementations are written in C++ and compiled using g++ v4.8.2 with the -O2 option.
We used the Mersenne Twister~\cite{matsumoto1998mersenne}, a general-purpose PRNG, to draw random numbers.

\begin{itemize}[leftmargin=*]
    \item \oneshot:
        In \textproc{Estimate},
        we invoke a PRNG to generate a random $x \in [0,1]$ for each examined edge $e$ and
        think of $e$ as alive if $ x < p(e) $.
        
    \item \snapshot:
        In \textproc{Build},
        we invoke a PRNG to generate a random $x \in [0,1]$ for each edge $e$ and each $i \in [r]$, and
        we keep $e$ in the $i$-th random graph $ G^{(i)} $ if $x < p(e)$.
    
    \item \ris:
        We use two kinds of PRNG,
        where one chooses a vertex in $V$ randomly, and
        the other generates a random real number between $[0,1]$.
        During a single RR-set generation in \textproc{Build},
        we first invoke the first kind of PRNG to generate a random target $v \in V$, and
        we then invoke the second kind of PRNG to generate a random $x \in [0,1]$ for each examined edge $e$ and
        think of $e$ as alive if $ x < p(e) $.
\end{itemize}

During greedy iterations in Algorithm~\ref{alg:review:framework},
two or more ``tie'' vertices may have the same maximum estimate.
To avoid dependency on the initially given order of vertices,
we shuffle the order randomly beforehand so that
simply running through vertices in this order breaks any ties randomly.

\subsection{Network Data}
Table~\ref{tab:design:dataset} summarizes the basic statistics of network data used in this paper.
Our selection includes not only real-world networks but also synthetic networks,
some of which are simple and small unlike those adopted in previous studies so that
we can run algorithms with a huge sample number.

\input{tab/design-dataset}

\subsubsection{Real-world Networks}

We use six social networks.
They are called \emph{complex networks}
sharing common structural features that occur in neither random graphs nor grid graphs.
\begin{description}
[leftmargin=\parindent]
\item[Scale-free property]
\cite{barabasi1999emergence}:
The distribution of vertex degrees follows power-law, i.e.,
the fraction of vertices having $k$ neighbors is  $\propto k^{-\gamma}$, where
$ \gamma \in [2, 3] $ typically.
\item[Small-world property]
\cite{watts1998collective}:
The expected distance between two vertices chosen randomly is small, typically $\bigO(\log n)$.
\item[Cluster property]
\cite{watts1998collective}:
Complex networks have a high clustering coefficient,
which is defined as 
three times the number of triangles over
the number of connected triplets.
\item[Core-whisker structure]
\cite{broder2000graph,leskovec2008statistical,maehara2014computing}:
Complex networks can be decomposed generally into two parts;
the expander-like dense ``core'' part and the tree-like ``whisker'' part.
\end{description}

\subsubsection{Synthetic Networks}
There exist plenty of generative models of random graphs aimed at resembling real networks' properties.
We adopt the \Barabasi-Albert model \cite{albert2002statistical}, which
generates scale-free undirected graphs by a preferential attachment process,
in which every time a new vertex has been added,
it is randomly connected to $M$ existing vertices.
In this paper,
we generate
a sparse version \BAs with $n = 1{,}000, M = 1$ and
a dense version \BAd with $n = 1{,}000, M = 11$, and
we assigned random directions for each edge.

\subsection{Edge Probabilities}
We finally designate edge probability settings.
Since publicly-available network data do not usually include influence probabilities,
we assign them artificially.
Our policy consists of the following well-established four strategies.

\begin{itemize}[leftmargin=*]
\item Uniform cascade (denoted \UC{0.1} (resp.~\UC{0.01})): 
    Each edge probability is a constant 0.1 (resp.~0.01).
\item In-degree weighted cascade (denoted \IWC):
    The influence probability of edge $(u,v)$ is set to $ 1/\ideg(v) $.
    It turns out that $ \sum_{u \in \inei(v)} p(u,v) = 1 $.
\item Out-degree weighted cascade (denoted \OWC):
    The influence probability of edge $(u,v)$ is set to $1/\odeg(u)$
    It turns out that $ \sum_{v \in \onei(u)} p(u,v) = 1 $.
\end{itemize}

Intuitively, each vertex owns equalized influence on its neighbors on \OWC while higher-degree vertices have more chances to activate their neighbors on \UC{}.

%% file: tab/design-dataset.tex
\begin{table}[tbp]
\caption{Network statistics. $\Delta^+$  and $\Delta^-$ denote maximum out- and in-degree.}
\label{tab:design:dataset}
\centering
\rowcolors{2}{gray!20}{white}
{\fontsize{7.5}{7.5}\selectfont
\setlength{\tabcolsep}{1pt}
\begin{tabular}{l|rrlrrccc} \toprule
\textbf{network} & $n$ & $m$ & \textbf{type} &
    $ \Delta^+ $ & $ \Delta^- $ & 
    \textbf{clus.~coef.} & 
    \textbf{avg.~dis.} \\
\midrule
& \multicolumn{7}{c}{\textit{real-world networks}} \\
\karate~\cite{kunegis2017konect} & 34 & 156 & social & 17 & 17 & 0.26 & 2.41 \\
\phy~\cite{kunegis2017konect} & 241 & 1,098 & social & 9 & 26 & 0.25 & -- \\
\grqc~\cite{leskovec2014snap} & 5,242 & 28,968 & collab. & 81 & 81 & 0.63 & -- \\
\wiki~\cite{leskovec2014snap} & 7,115 & 103,689 & voting & 893 & 457 & 0.13 & -- \\
\youtube & 1,134,889 & 5,975,248 & social & 28,754 & 28,754 & -- & -- \\
\pokec & 1,632,802 & 30,622,564 & social & 8,763 & 13,733 & -- & -- \\
\midrule
& \multicolumn{7}{c}{\textit{synthetic networks}} \\
\BAs & 1,000 & 999 & BA & 20 & 23 & 0.00 & 7.22 \\
\BAd & 1,000 & 10,879 & BA & 100 & 107 & 0.06 & 2.50 \\
\bottomrule
\end{tabular}
}
\end{table}

%% file: results.tex
\input{fig/H-karate}

\input{fig/H-tie}

\section{Results}
\label{sec:results}

In this section, we present the results of the experiments designed previously.
We conducted experiments on a Linux server with an Intel Xeon E5-2670 2.60GHz CPU and 500GB memory.
We constructed the empirical distribution of seed sets from $T = 1{,}000$ trials
for small instances and from $T = 20$ trials for large instances, which are marked with ``$\star$''.
Unless otherwise specified, the seed size $k$ was set to 1, 4, 16, 64, or 1,024, and
the sample number of
\oneshot ($\beta$), \snapshot ($\tau$), and \ris ($\theta$)
was set to a power of two up to
$2^{16}=65{,}536$,
$2^{16}=65{,}536$, and
$2^{24}=16{,}777{,}216$,
respectively,\footnote{Some of the results are missing when it took over weeks.
}
which results in
the convergence of seed set distributions
as shown in Section~\ref{sec:results:soldist}.
Our research focus is on the following.

\begin{itemize}[leftmargin=*]
    \item \textbf{Empirical distribution of seed sets} (Section~\ref{sec:results:soldist}):
    Since an influence estimator approaches to the exact influence function
    as the sample number increases,
    the seed set distribution is expected to settle down eventually.
    We will investigate how fast or slow the seed set distribution loses the diversity and verify its convergence for a variety of instances.
    We will also compare the diversity decay speed among \oneshot, \snapshot, and \ris.
    
    \item \textbf{Empirical distribution of influence spread} (Section~\ref{sec:results:infdist}):
    Similarly to the case of seed set distributions,
    it is expected that influence distributions converge eventually.
    We will observe how influence distributions concentrate, and
    assess the minimum sample number required to obtain near-optimal seed sets.
    We will further compare influence distributions among \oneshot, \snapshot, and \ris.
    For this purpose,
    we will discuss how to evaluate the \emph{quality} of influence distributions.

    \item \textbf{Empirical traversal cost} (Section~\ref{sec:results:cost}):
    Rather than merely measuring running time,
    which is dependent on the machine configuration,
    we employ the traversal cost.
    We will identify what kind of instance incurs high traversal cost.
    From Table~\ref{tab:review:taxonomy},
    the traversal-cost ratio of \oneshot, \snapshot, and \ris when $\beta=\tau=\theta$
    is that $1:1:\frac{1}{n}$ in theory;
    we will verify whether this is the case empirically.
\end{itemize}

\subsection{Distribution of Seed Sets}
\label{sec:results:soldist}

We first examine the distribution of seed sets.
Our task is to discover the speed at which the seed set distribution settles down.
To measure the diversity of a given distribution over sets,
we use the \emph{Shannon entropy},
$ H = - \sum_{S} p_S^{} \log_2 p_S^{} $,
where $ p_S^{} $ is the probability mass of vertex set $S$.
If the distribution takes only a single set, it is called \emph{degenerate} and has entropy 0.
Since the empirical distribution is constructed from $10^3$ trials,
its entropy never exceeds $\log_2 10^3 \approx 9.97$.

\subsubsection{Overall Tendencies}
We found that basically, the entropy in the early stages is nearly maximum, and it then monotonically decreases.
Figure~\ref{fig:soldist:H-karate} shows
the change in entropy as the sample number increases on \karate (\UC{0.1}, $k=1, 4, 16$).
In Figure~\ref{fig:soldist:H-karate:1},
where the maximum possible entropy is $ \log_2{34 \choose 1} \approx 5.09 $,
\oneshot reduced the entropy from 4.74 to 0,
\snapshot reduced the entropy from 4.92 to 0, and
\ris reduced the entropy from 5.06 to 0.
On instances including
\karate (\UC{0.1}, \UC{0.01}, \IWC, \OWC, $k=1$), \phy (\UC{0.01}, \OWC, $k=1$), \wiki (\UC{0.01}, \IWC, $k=1$), \BAs (\UC{0.1}, \UC{0.01}, \IWC, \OWC, $k=1$), and \BAd (\UC{0.01}, \IWC, $k=1$),
\BAs (\IWC, $k=16$),
$\star$ \youtube (\IWC, $k=1, 16$), and $\star$ \pokec (\IWC, $k=1, 4$),
the entropy for some algorithms eventually converged to 0.
We verified in those cases that the resulting seed sets are \emph{unique}
regardless of the choice of algorithm and sample number.
Therefore, the three algorithms have the \emph{same limit behavior}.
Figures~\ref{fig:soldist:H-karate:1} and~\ref{fig:soldist:H-karate:4} indicate that
for a fixed instance,
the entropy of algorithm $\alg$ at sample number $s$ can be represented by
$h(\alpha_\alg \cdot s)$ for some decreasing function $h$,
where $\alpha_\alg$ is a scaling parameter.
That is, \emph{the entropy of \oneshot, \snapshot, and \ris drops at the same rate up to scaling.}

\input{fig/H-ba}

\input{tab/inflist}
\input{fig/nbp-phy-u010}

\subsubsection{When is the Convergence Slow (or Fast)?}
We explain several cases where
the entropy slowly decreased or did not even converge to 0 (as far as we observed).
The main reason is that there are several ties so that we would need more samples to distinguish them from the limit.
On \karate (\IWC, $k=4$) and \phy (\IWC, $k=1$), shown in Figure~\ref{fig:soldist:H-tie},
the entropy stays around 1 for a long time, though eventually escapes from there.
This means that there are two almost-the-same-influence seed sets, and
the tie-breaking rule of Algorithm~\ref{alg:review:framework} selects either of them with almost equal probability.
In fact, the influence estimate for the two ties is
\hl{21.44}4 and \hl{21.44}6 on \karate (\IWC, $k=4$) and
\hl{12.4}03 and \hl{12.4}12 on \phy (\IWC, $k=1$),
which are nearly indistinguishable.
We also observed in Figure~\ref{fig:soldist:H-karate} that the larger the seed size $k$, the higher the entropy,
which is consistent with that the solution space is of size $ {n \choose k} $.

Here, we investigate the impact of edge probability settings on the entropy decay speed.
Figure~\ref{fig:soldist:H-ba} shows the result for the two \Barabasi-Albert networks when $k=1$.
\IWC shows the lowest entropy at any sample number on both networks while
\UC{0.01} and \OWC show the highest entropy on \BAs and \BAd, respectively.
Table~\ref{tab:soldist:inflist} reports
the top three influence spread of a single vertex for each instance.
One can see that
\emph{the large difference between 
the maximum influence and the second maximum 
is critical for quick convergence of entropy}.
Such differences in convergence speed tell us that experimental evaluation using a single setting of edge probabilities is not sufficient (e.g., \emph{only} \IWC was tested in \cite{tang2014influence,tang2015influence,nguyen2017social,nguyen2016stop,huang2017revisiting,nguyen2017billion,tang2018online}),
even though real data is unavailable.

\subsection{Distribution of Influence Spread}
\label{sec:results:infdist}

Let us now go into the distribution of influence spread.
Our task is to reveal
how fast the influence distribution concentrates, and
when the three algorithms yield the-same-quality influence distributions.
Since the exact computation of the influence spread is intractable, we used the following approximation scheme.
For each influence graph $\calG$, we generated $10^7$ RR sets, denoted $\calR_\calG$.
The unbiased estimator was then defined as
$ n \cdot F_{\calR_\calG}(\cdot) $.
This estimator was reused over different runs of different algorithms to ensure that
multiple identical seed sets have a unique estimate.
Note that the event that ``an RR set intersects $S$''
is a Bernoulli trial with success probability $ \frac{\Inf(S)}{n} $, and thus
the 99\% confidence interval for the actual influence spread is given by
$ n \cdot F_{\calR_\calG}(\cdot)  \pm \frac{1.29}{\sqrt{10^7}} n $.

\subsubsection{Overall Tendencies}

\begin{wrapfigure}[5]{r}{30mm}
\includegraphics[width=30mm,bb=10 150 580 540]{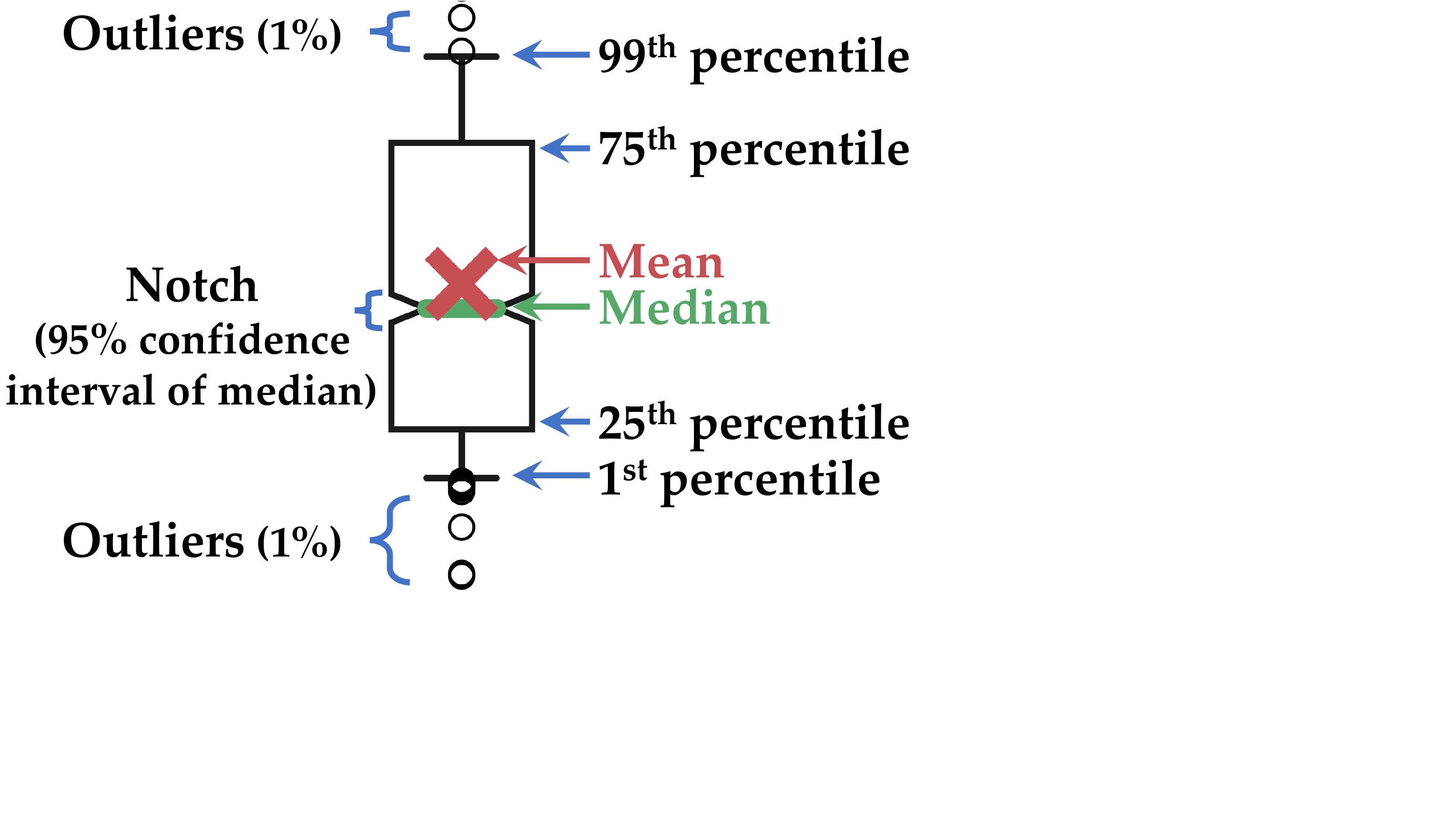}
\end{wrapfigure}

We visualize influence distributions.
Figure~\ref{fig:infdist:nbp-phy-u010}
shows notched box plots of influence distributions on
\phy (\UC{0.1}, $k=16$). See the right figure for  explanation of notched box plots.
Both mean and median are consistently increasing and (are expected to) eventually converge to the unique influence, which would be identical to Greedy on the exact influence.

\input{tab/near-optimal}

We then assess the sample number
needed to obtain ``near-optimal'' seed sets ``almost certainly.''
Specifically,
we regard the seed set uniquely obtained when $H = 0$ as \emph{Exact Greedy} and
a seed set of influence at least 0.95 times Exact Greedy as \emph{near-optimal}.
Table~\ref{tab:infdist:near-optimal} reports
the least sample number (denoted $\beta^*, \tau^*, \theta^*$) and
the respective entropy (denoted $H^*$) for which each algorithm was able to obtain a near-optimal seed set with probability at least 99\%
over $10^3$ trials.
The required sample number heavily depends on the problem instance, e.g.,
$\beta^*$ varies from 64 to 8,192 and $\tau^*$ varies from 16 to 8,192, and
\emph{it is therefore mandatory to appropriately select
the sample number of \oneshot and \snapshot} although the previous studies use a fixed sample number.
Remark also that the entropy does not have to be low, e.g., $H^*>4$ on \karate (\OWC, $k=4$).
We here show a large gap between
worst-case upper bounds and empirical least sample numbers:
On \wiki (\UC{0.01}, $k=4$) (resp.~\BAs (\IWC, $k=16$)),
the bound for \oneshot~\cite{tang2014influence} with $\epsilon=0.05, \delta=0.01$
is $1.0 \cdot 10^8 $ (resp.~$ 5.8 \cdot 10^7$)
while the empirical number is $256$ (resp.~$64$), and
the bound for \ris~\cite{tang2014influence} with $\epsilon=0.05, \delta=0.01$
is $ 1.6 \cdot 10^7 $ (resp.~$1.4 \cdot 10^6$)
and the empirical number is $131{,}072$ (resp.~$8{,}192$).

\subsubsection{Different Tendencies by Instances}
We discuss the difference by problem instances.
Reconsider Table~\ref{tab:infdist:near-optimal}
to examine the relation between the seed size $k$ and the required sample number.
In a theoretical analysis on stochastic submodular maximization~\cite{borgs2016maximizing,karimi2017stochastic},
the sample number linearly scales in $k$,
but this is not the case for our instances.
In particular,
on \karate (\UC{0.01}, \OWC) and \BAs (\IWC),
the required sample number decreases.
One possible reason is that there are numerous near-optimal seed sets;
a seed set that we extracted from them is reasonably influential even when the entropy is \emph{high}.
In fact, an \ris-type algorithm required
fewer RR sets when $k > 1$ than when $k = 1$,
e.g., see \cite[Figure 4d]{tang2015influence}.

\input{fig/nbp-grqc}

We then depict contrasting results from \grqc (\UC{0.1}, $k=1$) and \grqc (\OWC, $k=1$) in Figure~\ref{fig:infdist:nbp-grqc}.
In the former case,
the mean influence starts with less than 20\% of the maximum and
then quickly improves.
The 1st and 99th percentiles at $\theta = 2^{20}$ are 208.797 and 208.900, respectively.
In the latter case,
it slowly increases although the starting point is better than half of the maximum, and
the 1st and 99th percentiles at $\theta = 2^{20}$ are 6.795 and 7.429, respectively.
The core-whisker structure can best explain this.
In the random graph counterpart of \grqc (\UC{0.1}),
a certain portion of the densely-connected core part forms a \emph{giant component}~\cite{karp1990transitive,bollobas2001random,ohsaka2017coarsening}, and
the tree-like whisker part becomes entirely disconnected.
Hence, there exist the two extremes of very influential vertices and poorly influential vertices, and
it becomes easier to identify some core vertices as soon as the sample number grows.
On the other hand,
in the random graph counterpart of \grqc (\OWC),
all vertices have exactly one outgoing edge in expectation, and
so are similarly influential;
however, it is difficult to distinguish
the best influential vertex from a number of \emph{slightly less} influential ones.

\subsubsection{Comparison among Algorithms}
We present the comparison among \oneshot, \snapshot, and \ris.
We first discuss how to measure the \emph{quality} of influence distributions.
One might think that we need to evaluate several statistics such as
mean, standard deviation, and percentiles simultaneously.
Our important finding is that for each instance,
the \emph{mean} can be a dominant factor.
In Figure~\ref{fig:infdist:std} (resp.~\ref{fig:infdist:skewness}),
we demonstrate that the relation between mean and standard deviation (resp.~first percentile)
is almost independent of the choice among \oneshot, \snapshot, and \ris.
We thus declare that influence distribution $\calI_1$ is
\emph{better than} influence distribution $\calI_2$ if
the mean of $\calI_1$ is greater than that of $\calI_2$.

\input{fig/mean_vs_std}

We then determine what sample number and sample size make two distributions comparable with each other as follows.
Fix an instance.
For each $i \in \{1,2\}$,
let $\alg_i$ be one of \oneshot, \snapshot, and \ris,
$s_i$ denote the sample number of $\alg_i$, and
$\calI_i^{(s_i)}$ be the distribution of influence spread obtained by running $\alg_i$ with $s_i$.
Then,
$s_2$ is said to be \emph{comparable} to $s_1$ if
``$s_2$ is the least sample number such that $\calI_2^{(s_2)}$ is better than $\calI_1^{(s_1)}$,''
$\sfrac{s_2}{s_1}$ is called the \emph{comparable number ratio} of $\alg_2$ to $\alg_1$, and
$ (\text{sample size of } \alg_2 \text{ at } s_2) / (\text{sample size of } \alg_1 \text{ at } s_1) $ is called the \emph{comparable size ratio} of $\alg_2$ to $\alg_1$.
A large ``number'' ratio means that
$\alg_2$ requires more samples than $\alg_1$
while a large ``size'' ratio means that
$\alg_2$ requires larger samples in size than $\alg_1$.
Note that $\sfrac{s_2}{s_1}$ takes a power of two since so do $s_1$ and $s_2$.

\paragraph*{\oneshot versus \snapshot}
We first investigate the comparable number ratio of \oneshot to \snapshot as shown in Figure~\ref{fig:infdist:cr-ss-os}.\footnote{
Since the sample size of \oneshot is 0, the size ratio is not analyzed.
}
The comparable ratio almost always lies within the range from 1 to 32; i.e.,
\snapshot requires fewer samples than \oneshot.
Besides, the ratio is stable with \snapshot's sample number $\tau$, that is,
\emph{the mean influence of \oneshot and \snapshot improves at the same rate up to scaling of sample number}.
We thus report the median of the comparable number ratio in Table~\ref{tab:infdist:median-cr-ss-os}, which increases to up to 96 as the seed size $k$ increases and further study might be needed.

\input{fig/cr-ss-os}
\input{tab/median-cr-ss-os}

\paragraph*{\snapshot versus \ris}

We next investigate both the comparable size ratio and comparable number ratio
of \ris to \snapshot shown in
Figure~\ref{fig:infdist:cr-ss-ris} and Table~\ref{tab:infdist:median-cr-ss-ris}.
\emph{The comparable size ratio is stable with
\snapshot's sample size $\tau \cdot \tilde{m}$}, and
this is the case for the comparable number ratio.
The comparable number ratio varies from $2^2 = 4$ to up to $2^{19} = 524{,}288$, and
it is less dependent on seed size $k$ unlike the comparison between \oneshot and \snapshot.

Comparable number ratios over 4,096 are observed on
\phy (\UC{0.01}), \grqc (\UC{0.01}), \youtube (\UC{0.01}, \IWC, \OWC),
\pokec (\UC{0.01}, \OWC), and \BAs (\UC{0.01}).
The primary reason is that the influence spread on these instances is tiny and so is an RR set;
the upper bound on \ris's sample number $\theta$~\cite{borgs2016maximizing} is inversely proportional to the maximum influence (divided by $n$).
We emphasize that \emph{just because \snapshot requires much fewer samples than \ris
does not mean that \snapshot is more efficient than \ris.}
Comparable size ratios have a quite different trend, which are significantly less than 1,
e.g., 0.00033 on \youtube (\IWC, $k=1$) and 0.016 on \pokec (\IWC, $k=1, 4$).
We thus conclude that
when \ris is comparable to \snapshot,
\emph{\snapshot requires fewer but larger samples than \ris},
or equivalently,
\emph{\ris is more space-saving than \snapshot}
for large networks.

\input{fig/cr-ss-ris}
\input{tab/median-cr-ss-ris}

\input{tab/results-cost}

\subsection{Graph Traversal Cost per Sample}
\label{sec:results:cost}

We finally examine the per-sample traversal cost of naive implementations.
Table~\ref{tab:cost} reports
the average traversal cost corresponding to the case of $k=1$ and $\beta=\tau=\theta=1$.

\subsubsection{Overall Tendencies}
The edge traversal cost is higher than the vertex traversal cost excepting \snapshot on several instances.
Generally, the larger the influence graph, the higher the  traversal cost.
The setting of edge probabilities also has a significant impact;
\UC{0.1} often incurs the most expensive cost, e.g.,
\BAd (\UC{0.1}) and \grqc (\UC{0.1}). 
This can be explained by the core-whisker structure:
A giant component is likely to appear
in the random graph counterpart of \BAd (\UC{0.1})
but not of \BAd (\UC{0.01}, \IWC, and \OWC),
whose expected number of edges $\tilde{m}$ is less than $n$.
Such an influence graph that a giant component is likely to appear
will be referred to as a \emph{giant-component influence graph}.

\subsubsection{Comparison among Algorithms}
Firstly, we compare \oneshot with \snapshot.
The vertex traversal cost is the same as expected from Table~\ref{tab:review:taxonomy},
but the edge traversal cost is different.
Specifically,
the edge traversal cost of \snapshot is approximately $ \tilde{m}/m $ times that of \oneshot, where
$\tilde{m}$ is $0.1m$ on \UC{0.1}, $0.01m$ on \UC{0.01}, and $n$ on \IWC and \OWC.
This factor comes from that \snapshot scans only live edges in \textproc{Estimate}.

We then compare \oneshot with \ris.
The vertex traversal cost of \oneshot is $n$ times that of \ris as expected from Table~\ref{tab:review:taxonomy}.
The edge traversal cost indicates a similar tendency,
e.g., the cost of \oneshot divided by that of \ris is within
[240, 255] on \phy with 241 vertices and
[616, 1,615] on \BAd with 1,000 vertices,
which is not implied by Table~\ref{tab:review:taxonomy}.
It seems that there is not much difference between an influence graph and its transposed counterpart
with regard to the edge traversal cost
though $ \Inf_{\calG} \neq \Inf_{\calG^\top} $.
To sum up, \emph{there is a simple edge-traversal-cost relation: $ \oneshot \approx m/\tilde{m} \cdot \snapshot \approx n \cdot \ris $, and \ris is the most per-sample time-efficient}.

\subsection{Recap of Findings}
\label{sec:results:summary}

So far, we have tested
naive implementations of \oneshot, \snapshot, and \ris
on eight graph data under four edge probability settings.
Our empirical findings are summarized below, and
the next section is devoted to further discussions.

\subsubsection{Distribution of Seed Sets}
Seed set distributions approach to degenerate as the sample number grows, i.e.,
for a sufficiently large sample number, we obtain a single seed set.
This seed set is unique regardless of the choice of algorithms, which means that
\oneshot, \snapshot, and \ris have the same limit behavior.
The entropy of \oneshot, \snapshot, and \ris drops
at the same rate up to scaling of sample number.
However, the entropy slowly decreases
if several ties have almost-the-same influence.
In particular, the large difference between the maximum influence and the second maximum is critical for quick convergence.

\subsubsection{Distribution of Influence Spread}
Following the convergence nature of seed set distributions,
influence distributions converge to the unique influence.
In particular, the mean influence monotonically increases as the sample number increases.
Concerning the tail behavior,
the minimum sample number required to obtain near-optimal seed sets
takes a wide range of values, i.e.,
$\beta^* \in [64, 8{,}192]$,
$\tau^* \in [16, 8{,}192]$, and
$\theta^* \in [4{,}096, 1{,}048{,}576]$,
depending on edge probability settings, seed size, and graph size.
Hence, it is mandatory for \oneshot- and \snapshot-type algorithms to appropriately select the sample number.
Comparing among the three approaches,
the mean influence of \oneshot, \snapshot, and \ris improves
at the same rate up to scaling of sample number and sample size.
\oneshot requires up to 96 times as many samples as \snapshot requires,
\snapshot requires fewer (say, $10^5$ times fewer) but larger (say, $10^3$ times larger) samples in size than RIS;
i.e., RIS is more space-saving than Snapshot.

\subsubsection{Traversal Cost}
High traversal costs are incurred on a giant-component influence graph
such as \grqc (\UC{0.1}) and \BAd (\UC{0.1})
having both high-degree vertices and high-probability edges.
The empirically observed relation among \oneshot, \snapshot, and \ris is that
$ 1:1:\frac{1}{n} $ for vertex traversal cost and
$ 1:\frac{\tilde{m}}{m}:\frac{1}{n}$ for edge traversal cost,
demonstrating that
\ris offers the best per-sample time-efficiency and
\snapshot offers the second best.

%% file: fig/H-karate.tex
\begin{figure*}[htbp]
\centering
\subfloat[\karate (\UC{0.1}, $k=1$)\label{fig:soldist:H-karate:1}]{
	\includegraphics[width=0.3\hsize]{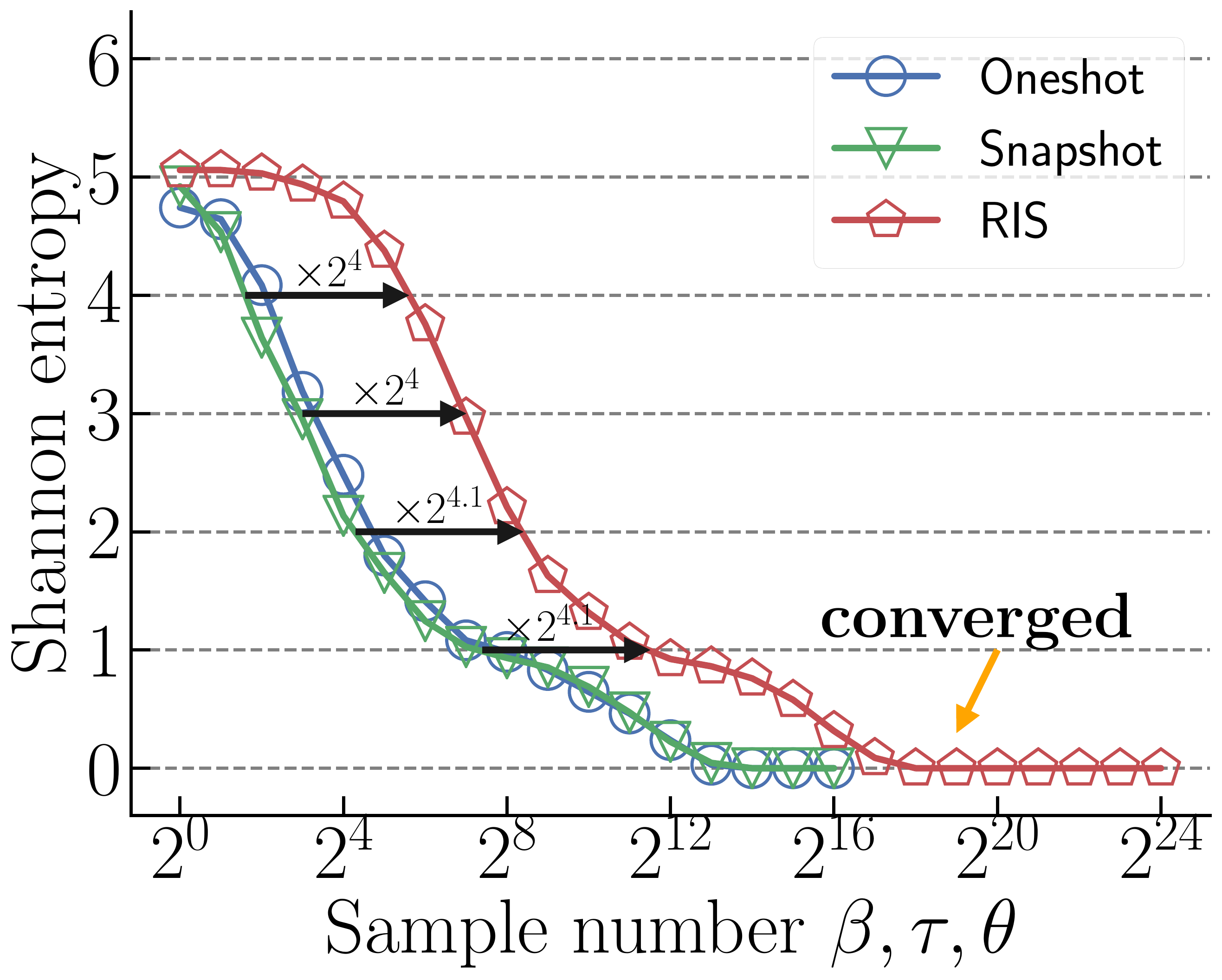}
}
\subfloat[\karate (\UC{0.1}, $k=4$)\label{fig:soldist:H-karate:4}]{
	\includegraphics[width=0.3\hsize]{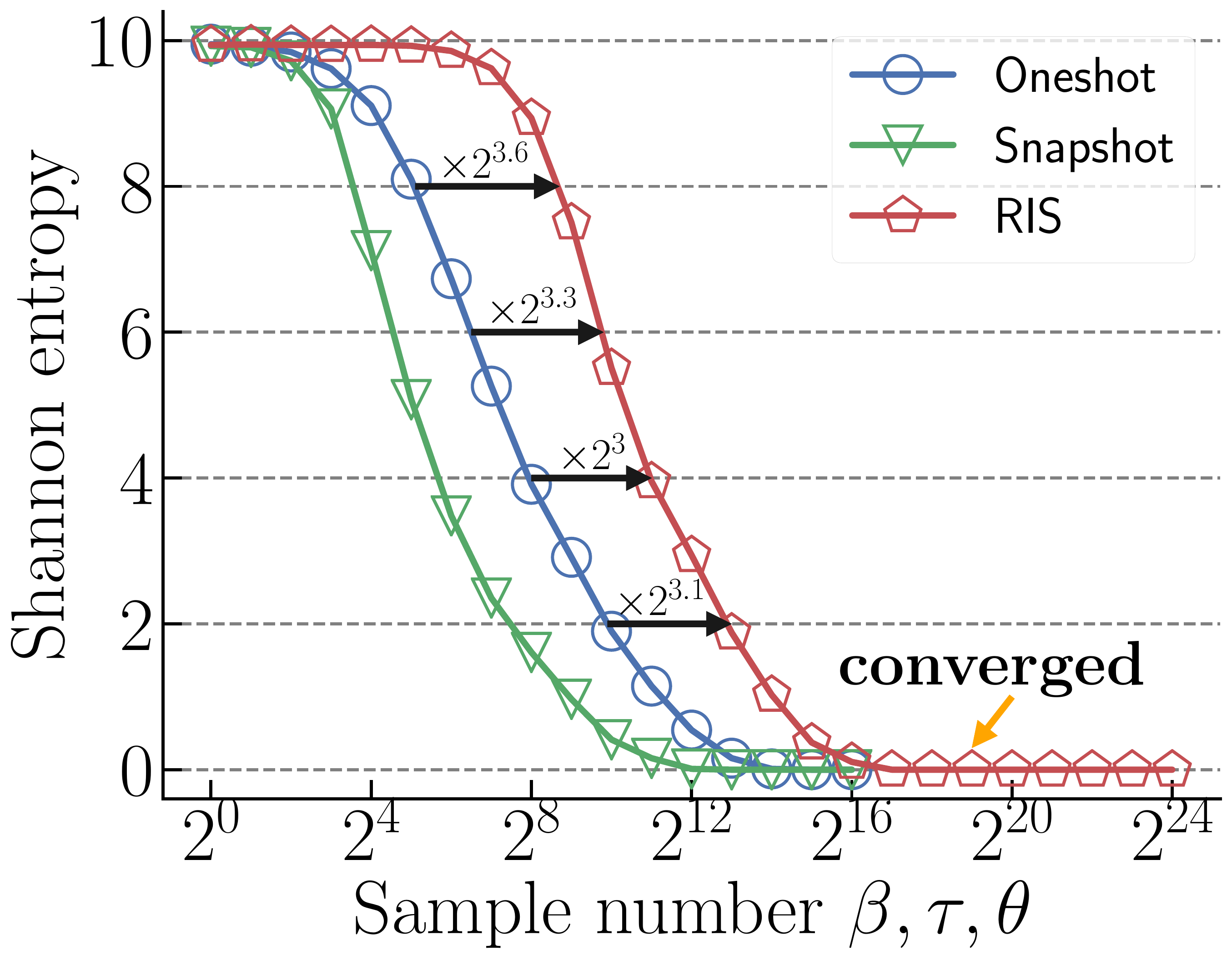}
}
\subfloat[\karate (\UC{0.1}, $k=16$)\label{fig:soldist:H-karate:16}]{
	\includegraphics[width=0.3\hsize]{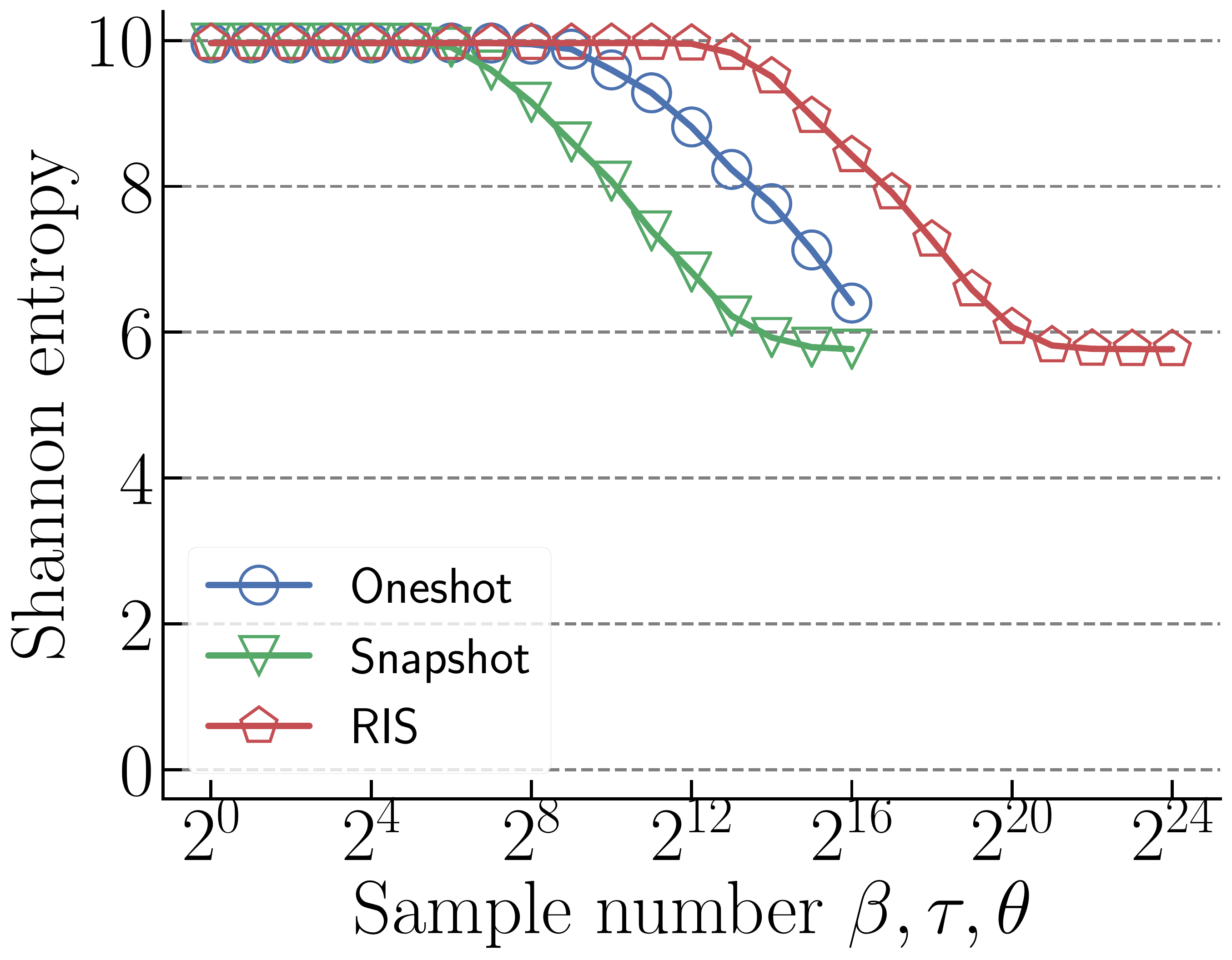}
}
\caption{Entropy of the seed set distributions on \karate (\UC{0.1}).
The entropy converged to 0 when $k=1, 4$.}
\label{fig:soldist:H-karate}
\end{figure*}

%% file: fig/H-tie.tex
\begin{figure}[tbp]
\centering
\subfloat[\karate (\IWC, $k=4$)]{
	\includegraphics[width=0.5\hsize]{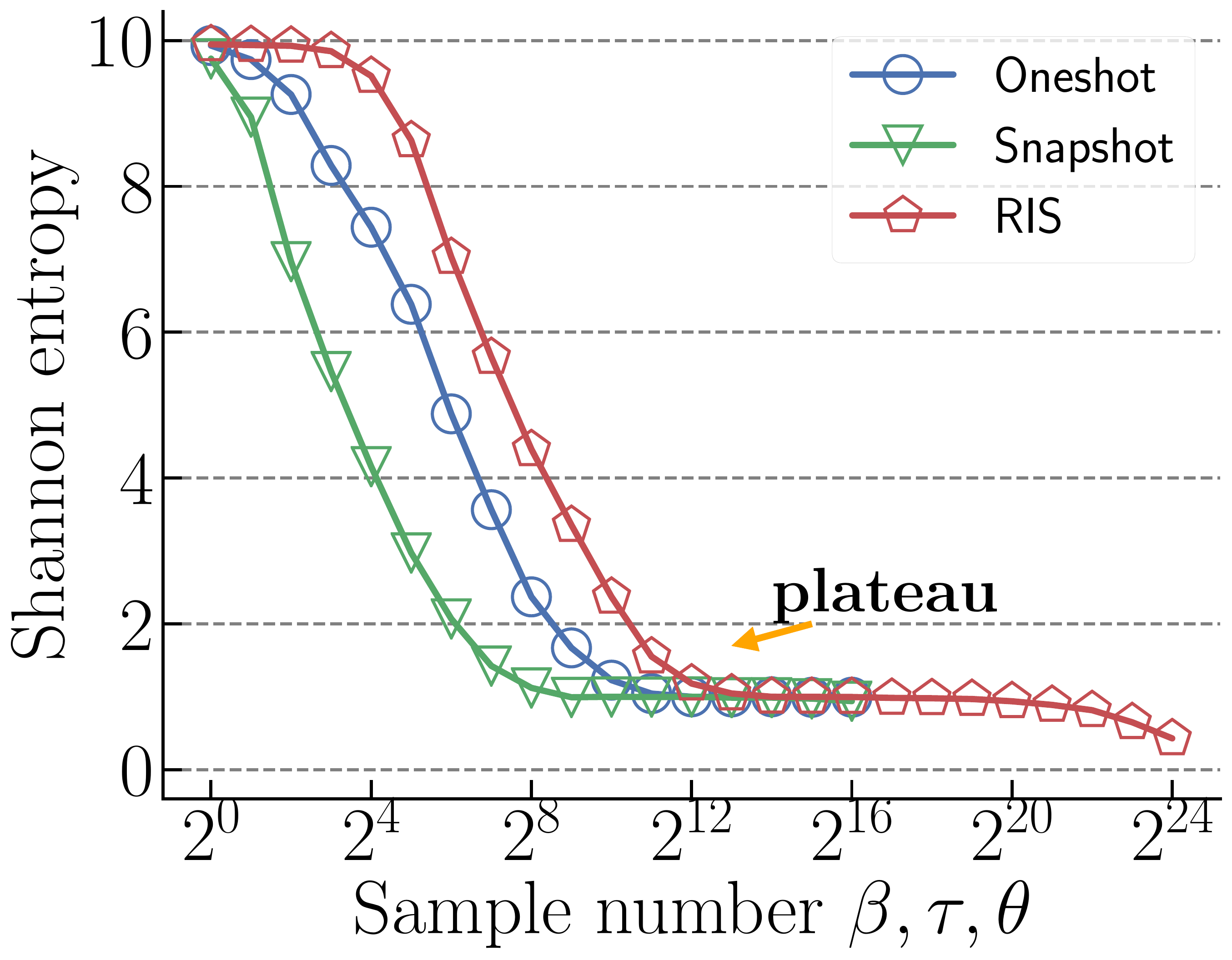}
}
\subfloat[\phy (\IWC, $k=1$)]{
	\includegraphics[width=0.5\hsize]{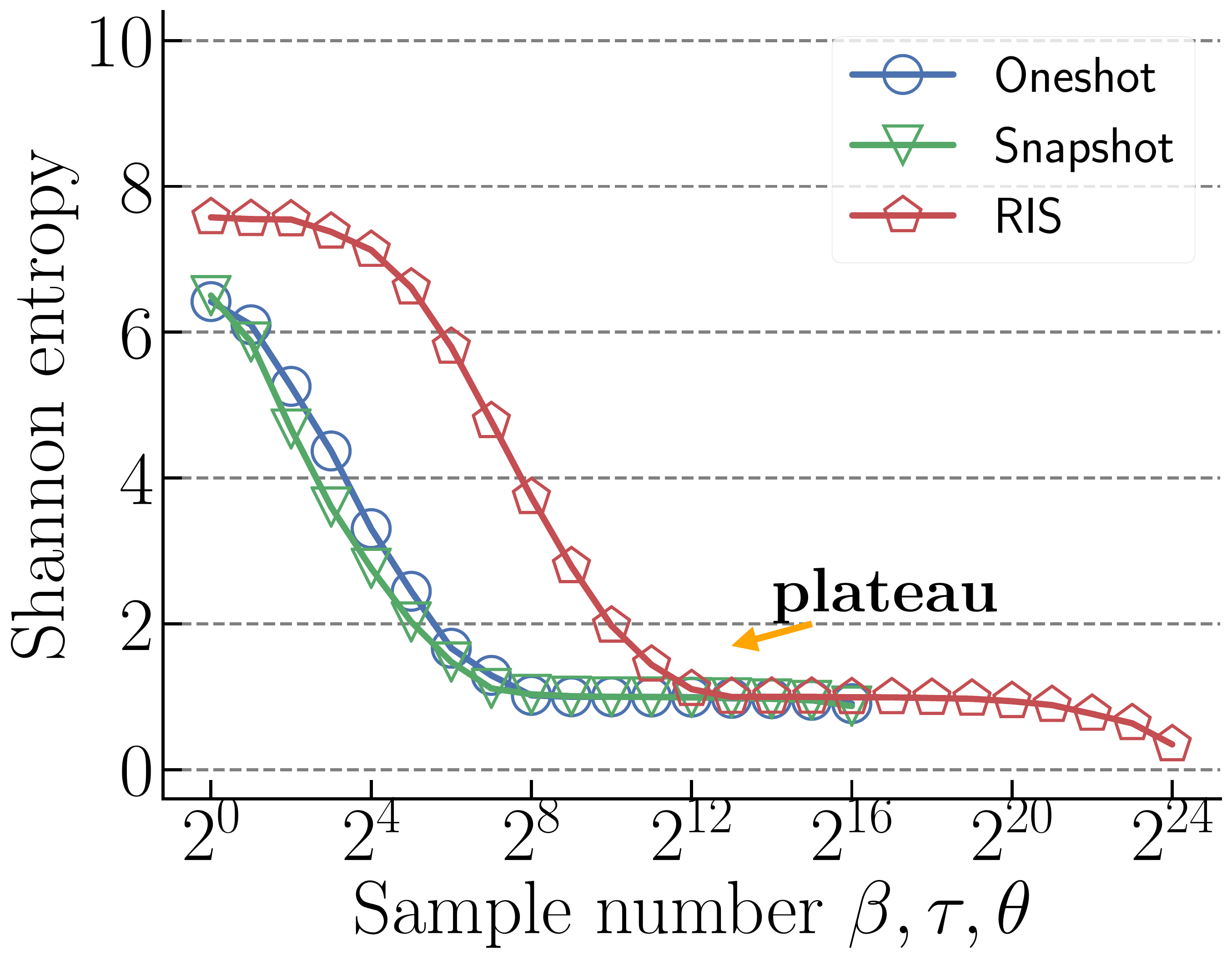}
}
\caption{Two instances for which the entropy hits a plateau,
which means that there exist two almost-the-same-influence seed sets.}
\label{fig:soldist:H-tie}
\end{figure}

%% file: fig/H-ba.tex
\begin{figure}[tbp]
\centering
\subfloat[\BAs, $k=1$, \ris\label{fig:soldist:H-ba_s}]{
	\includegraphics[width=0.5\hsize]{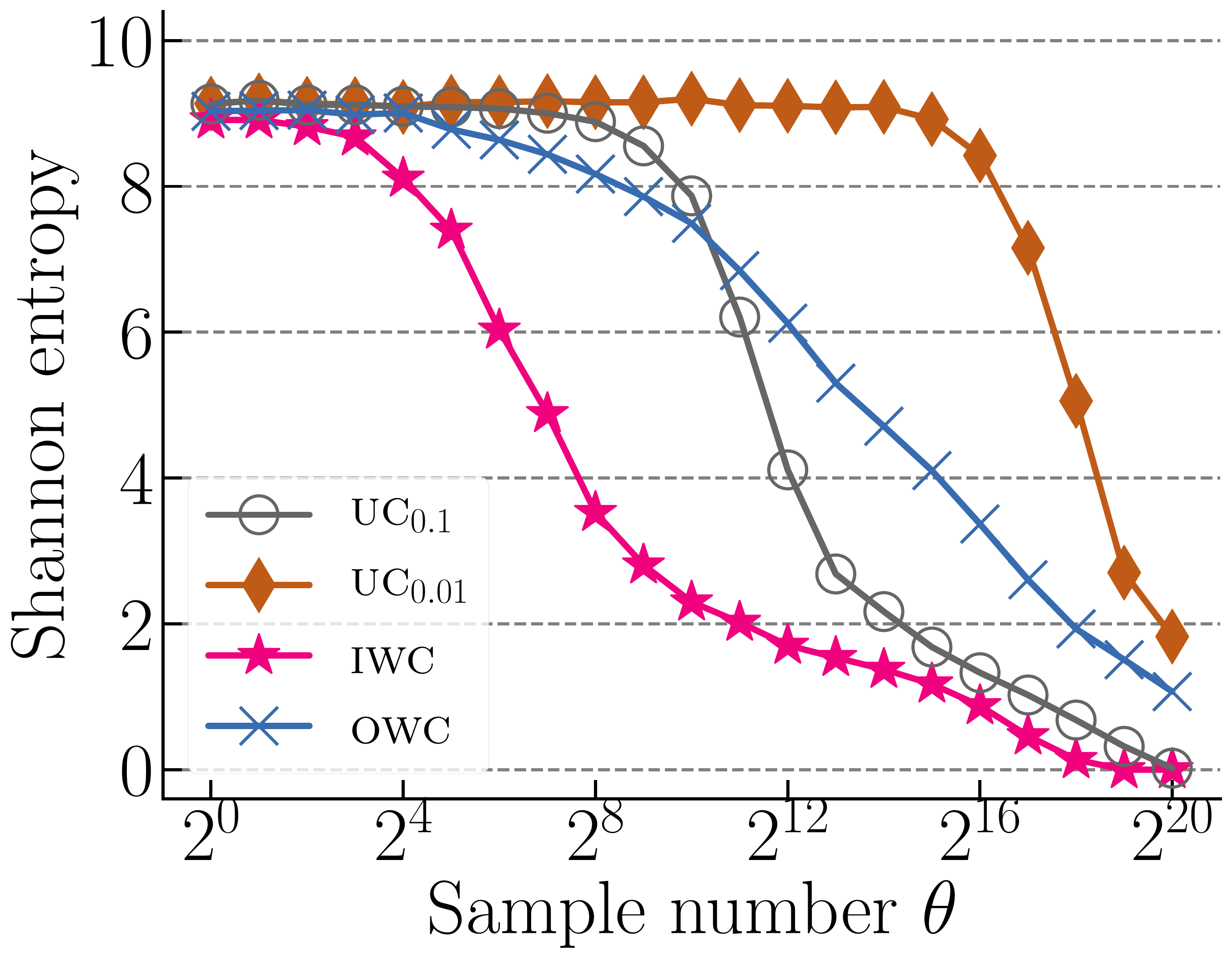}
}
\subfloat[\BAd, $k=1$, \ris\label{fig:soldist:H-ba_d}]{
	\includegraphics[width=0.5\hsize]{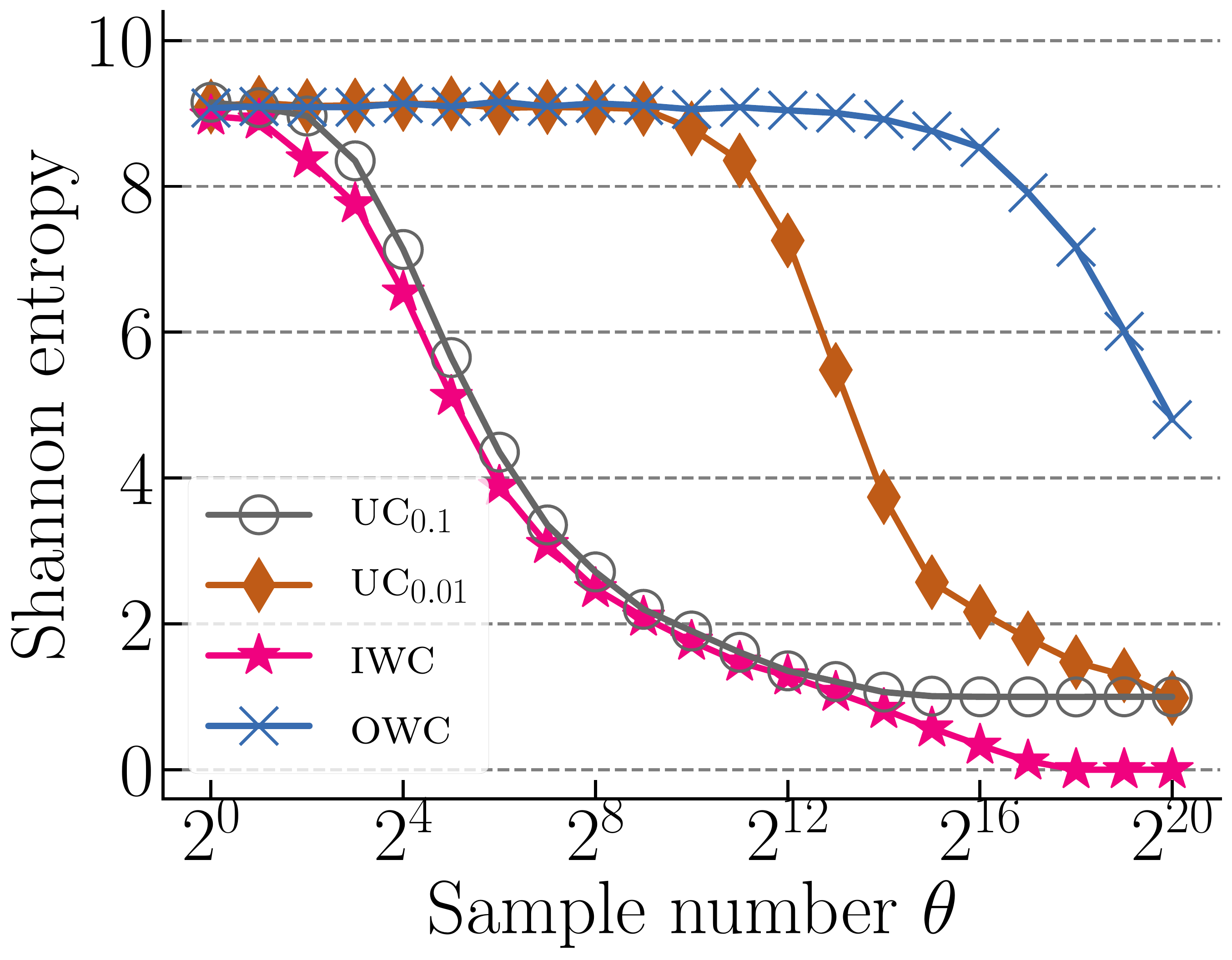}
}
\caption{Difference of the entropy decay speed by edge probability settings.}
\label{fig:soldist:H-ba}
\end{figure}

%% file: tab/inflist.tex
\begin{table}[tbp]
    \centering
    \caption{Top three influence spread of a single vertex.
    \hl{Highlighted digits} match to those of $\Inf(v^\mathrm{1st})$.
    }
    \label{tab:soldist:inflist}
\rowcolors{2}{gray!20}{white}
{\fontsize{7.5}{7.5}\selectfont
\begin{tabular}{c|rrrr}
\toprule
\BAs (Figure~\ref{fig:soldist:H-ba_s}) &
\color[HTML]{666666}{\UC{0.1}} &
\color[HTML]{bf5b17}{\UC{0.01}} &
\color[HTML]{f0027f}{\IWC} &
\color[HTML]{386cb0}{\OWC} \\
$\Inf(v^\mathrm{1st})$ & \hl{3.2089} & \hl{1.1901} & \hl{21.4167} & \hl{5.1365} \\
$\Inf(v^\mathrm{2nd})$ & 2.9921 & \hl{1.1}739 & \hl{2}0.5095 & \hl{5}.0286 \\
$\Inf(v^\mathrm{3rd})$ & 2.9779 & \hl{1.1}604 & \hl{2}0.3244 & \hl{5}.0272 \\
\midrule
\BAd (Figure~\ref{fig:soldist:H-ba_d}) &
\color[HTML]{666666}{\UC{0.1}} &
\color[HTML]{bf5b17}{\UC{0.01}} &
\color[HTML]{f0027f}{\IWC} &
\color[HTML]{386cb0}{\OWC} \\
$\Inf(v^\mathrm{1st})$ & \hl{377.0686} & \hl{2.1710} & \hl{101.7954} & \hl{15.5098} \\
$\Inf(v^\mathrm{2nd})$ & \hl{377.0}483 & \hl{2.1}162 & \hl{10}0.1006 & \hl{15.50}31 \\
$\Inf(v^\mathrm{3rd})$ & \hl{37}5.7611 & \hl{2.1}066 & 96.1872 & \hl{15}.4971 \\
\bottomrule
\end{tabular}
}
\end{table}

%% file: fig/nbp-phy-u010.tex
\begin{figure*}[tbp]
\centering
\subfloat[\oneshot]{\includegraphics[width=0.28\hsize]{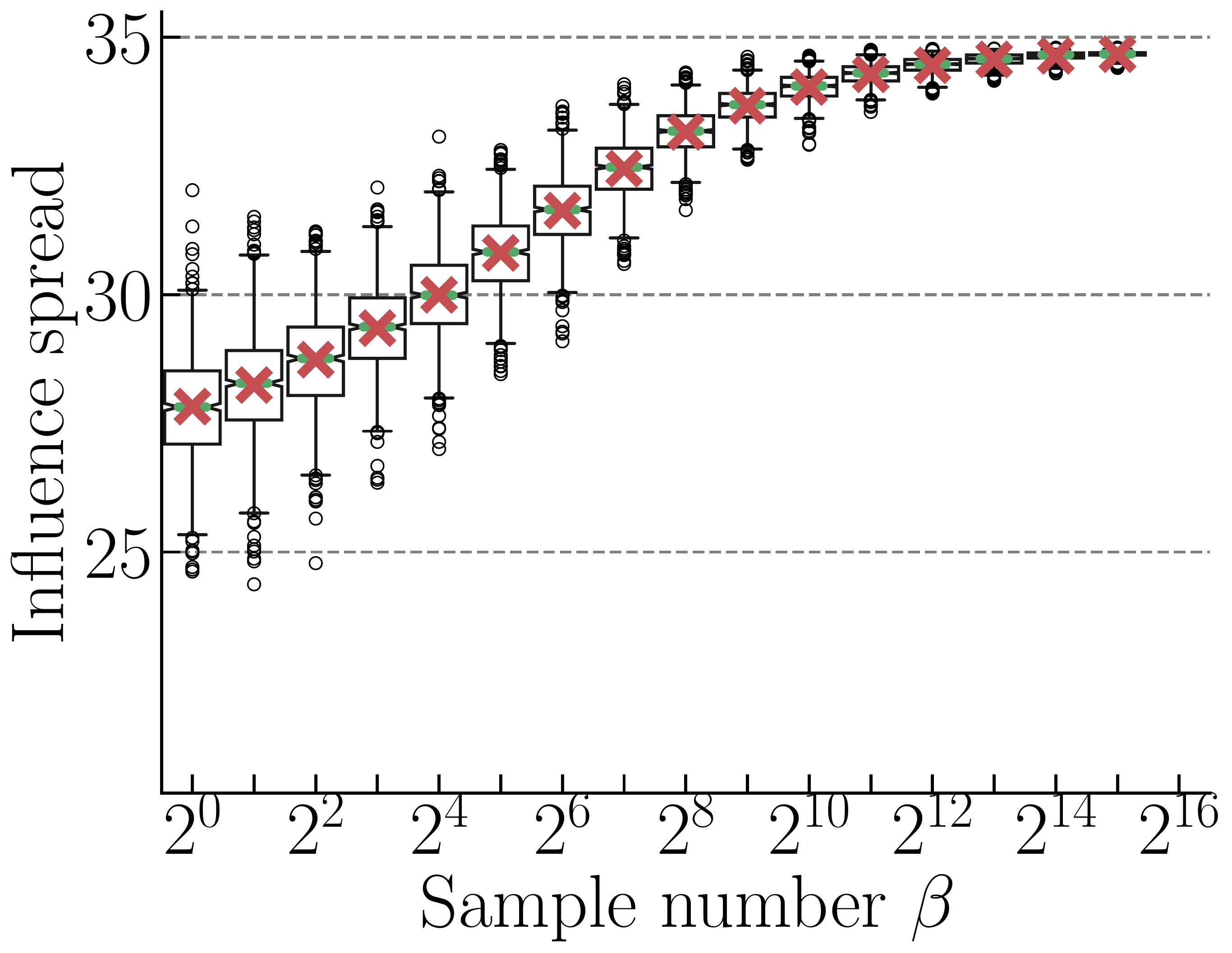}}
\subfloat[\snapshot]{\includegraphics[width=0.28\hsize]{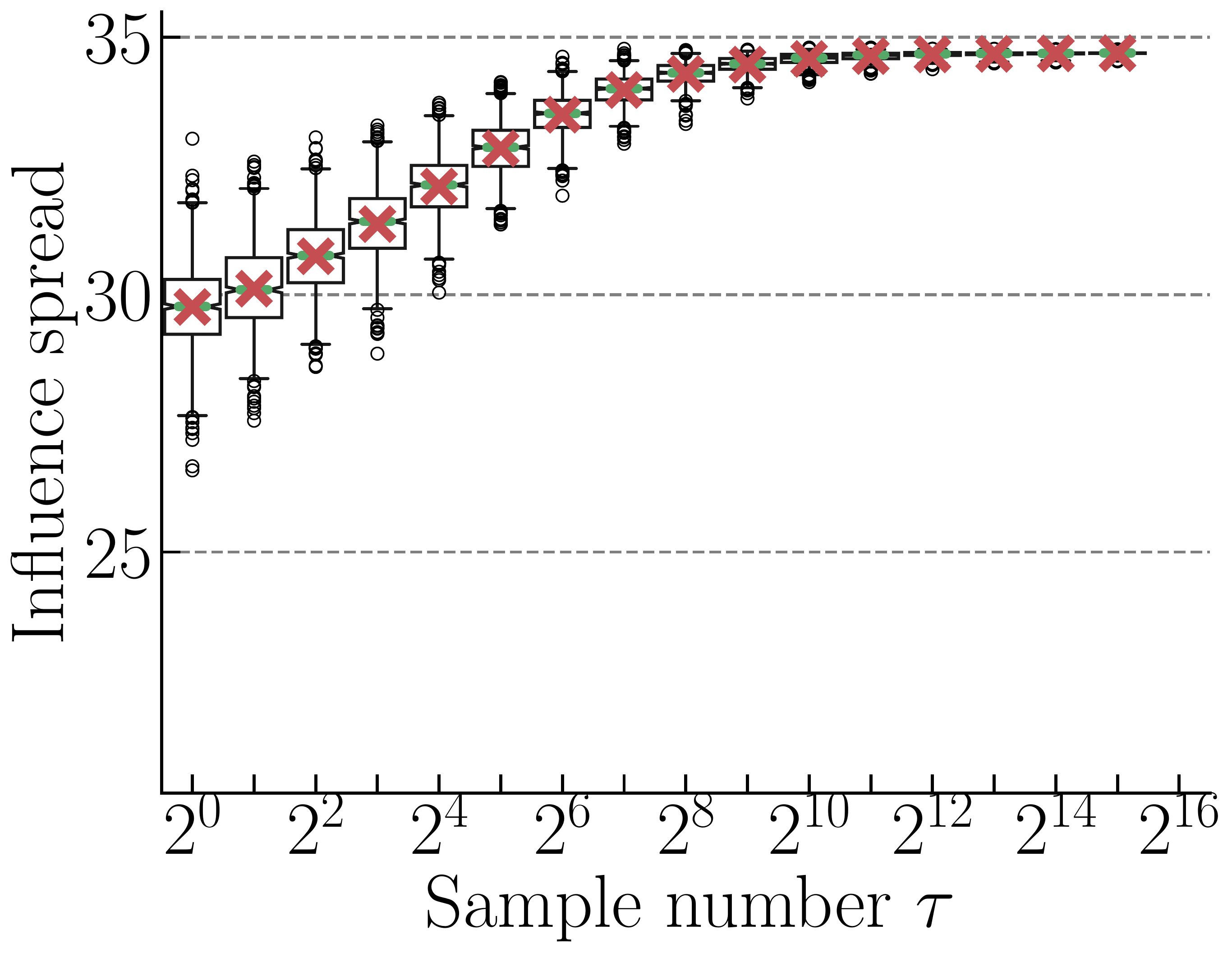}}
\subfloat[\ris]{\includegraphics[width=0.28\hsize]{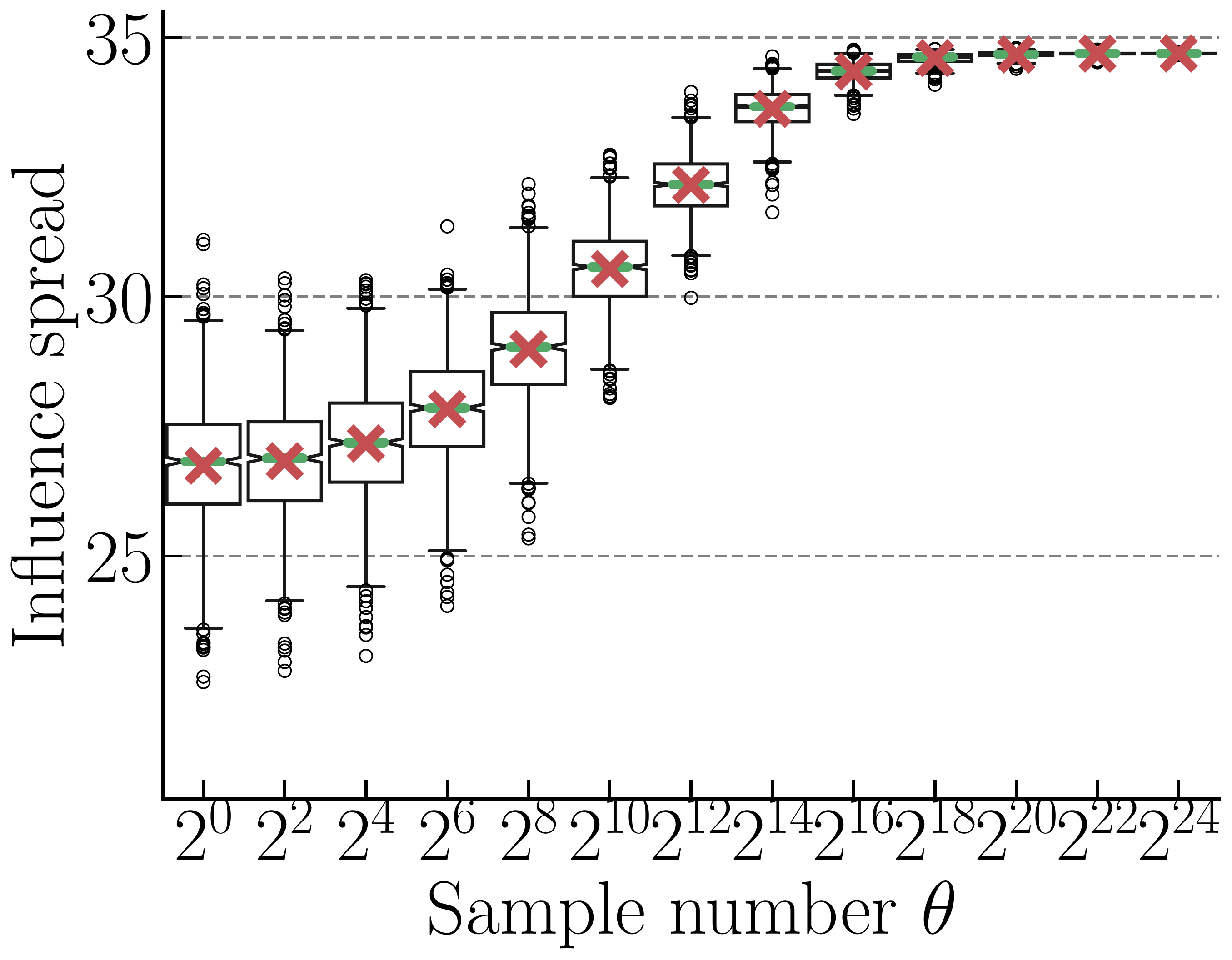}}
\\
\caption{Influence distribution in notched box plot on \phy (\UC{0.1}, $k=16$).}
\label{fig:infdist:nbp-phy-u010}
\end{figure*}

%% file: tab/near-optimal.tex
\begin{table}[tbp]
    \centering
    \caption{Least sample number and 
    corresponding entropy required to obtain near-optimal seed sets with probability 99\% for each algorithm.
    \textcolor{Blue3}{Blues} and \textcolor{Red3}{reds} are for minimum and maximum, respectively.}
    \label{tab:infdist:near-optimal}
\rowcolors{2}{gray!20}{white}
{\fontsize{7.5}{7.5}\selectfont
\setlength{\tabcolsep}{3pt}
\begin{tabular}{llr|rr|rr|rr}
\toprule
& & & \multicolumn{2}{c|}{\oneshot}
& \multicolumn{2}{c|}{\snapshot}
& \multicolumn{2}{c}{\ris} \\
\textbf{network} & \textbf{prob.} & $k$ &
$\log_2 \beta^*$ & $H^*$ &
$\log_2 \tau^*$ & $H^*$ &
$\log_2 \theta^*$ & $H^*$ \\
\midrule
\karate &   \UC{0.1} &  1 &  8 &  0.98 &   7 &  1.03 &  \mini{12} &  0.92 \\
\karate &   \UC{0.1} &  4 &  9 &  2.91 &   7 &  2.35 &  \mini{12} &  2.94 \\
\karate &  \UC{0.01} &  1 &  8 &  1.14 &   8 &  1.13 &         16 &  1.14 \\
\karate &  \UC{0.01} &  4 &  7 &  7.10 &   7 &  3.43 &         15 &  5.46 \\
\karate &       \IWC &  1 & 10 &  0.06 &  10 &  0.04 &         14 &  0.02 \\
\karate &       \OWC &  1 & 11 &  0.35 &  11 &  0.24 &         14 &  0.46 \\
\karate &       \OWC &  4 & 10 &  4.49 &   8 &  5.57 &  \mini{12} &  5.67 \\
\phy & \UC{0.01} & 1 &          7 &  5.95 &          7 &  5.90 &  \maxi{20} &  5.15 \\
\phy &      \IWC & 4 &         10 &  2.15 &          8 &  1.80 &         13 &  2.67 \\
\phy &      \OWC & 1 &  \maxi{13} &  0.23 &  \maxi{13} &  0.14 &         17 &  0.45 \\
\wiki & \UC{0.01} &  1 &  7 &  0.92 &  7 &  0.99 &  18 &  0.83 \\
\wiki & \UC{0.01} &  4 &  7 &  1.79 &  6 &  1.26 &  17 &  1.26 \\
\wiki &      \IWC &  1 &  7 &  0.25 &  7 &  0.22 &  17 &  0.09 \\
\wiki &      \IWC &  4 &  7 &  0.87 &  5 &  0.77 &  15 &  0.56 \\
\wiki & \UC{0.01} &  1 &  8 &  0.62 &  8 &  0.61 &  19 &  0.51  \\
\wiki & \UC{0.01} &  4 &  8 &  0.78 &  6 &  1.26 &  17 &  1.26  \\
\wiki &      \IWC &  1 &  9 &  0.01 &  8 &  0.07 &  18 &  0.01 \\
\wiki &      \IWC &  4 &  8 &  0.37 &  6 &  0.23 &  15 &  0.56 \\
\BAs &  \UC{0.1} &  1 &        10 &  0.03 &        10 &  0.02 &      \maxi{20} &  0.02 \\
\BAs & \UC{0.01} &  1 &         8 &  1.32 &         9 &  0.97 &  \maxi{$>$ 20} & -- \\
\BAs &      \IWC &  1 &         9 &  0.08 &        10 &  0.01 &             18 &  0.14 \\
\BAs &      \IWC & 16 &  \mini{6} &  8.18 &  \mini{4} &  5.10 &             13 &  7.52 \\
\BAs &      \OWC &  1 &         9 &  0.07 &         9 &  0.08 &             19 &  1.51 \\
\BAd & \UC{0.01} &  1 &         8 &  1.35 &         8 &  1.42 &             18 &  1.47 \\
\BAd &      \IWC &  1 &        11 &  0.78 &        11 &  0.74 &             14 &  0.83 \\
\bottomrule
\end{tabular}
}
\end{table}

%% file: fig/nbp-grqc.tex
\begin{figure}[tbp]
\captionsetup[subfigure]{labelformat=empty}
\centering
\subfloat[Quick convergence on \UC{0.1}]{\includegraphics[width=0.5\hsize]{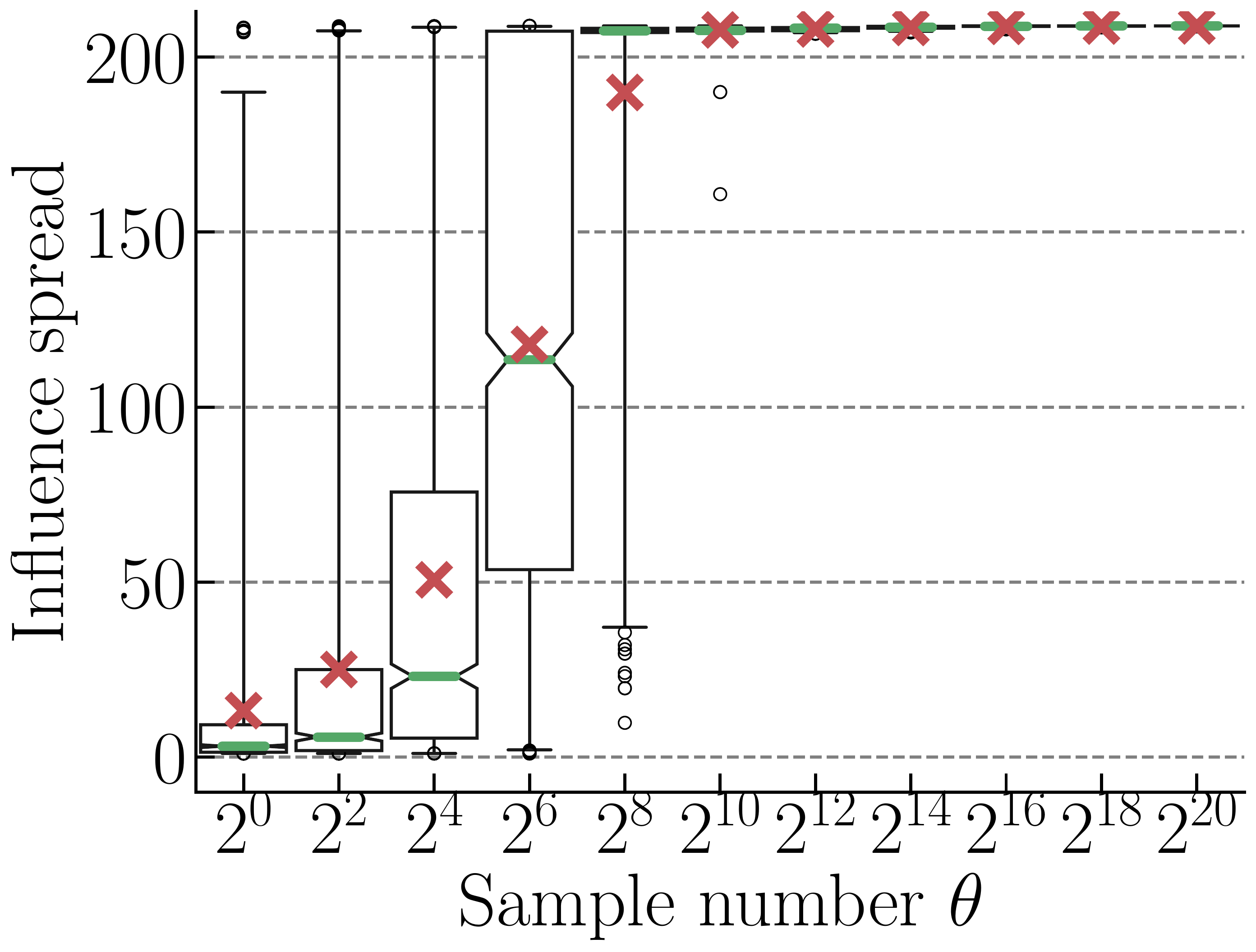}}
\subfloat[Slow improvement on \OWC]{\includegraphics[width=0.5\hsize]{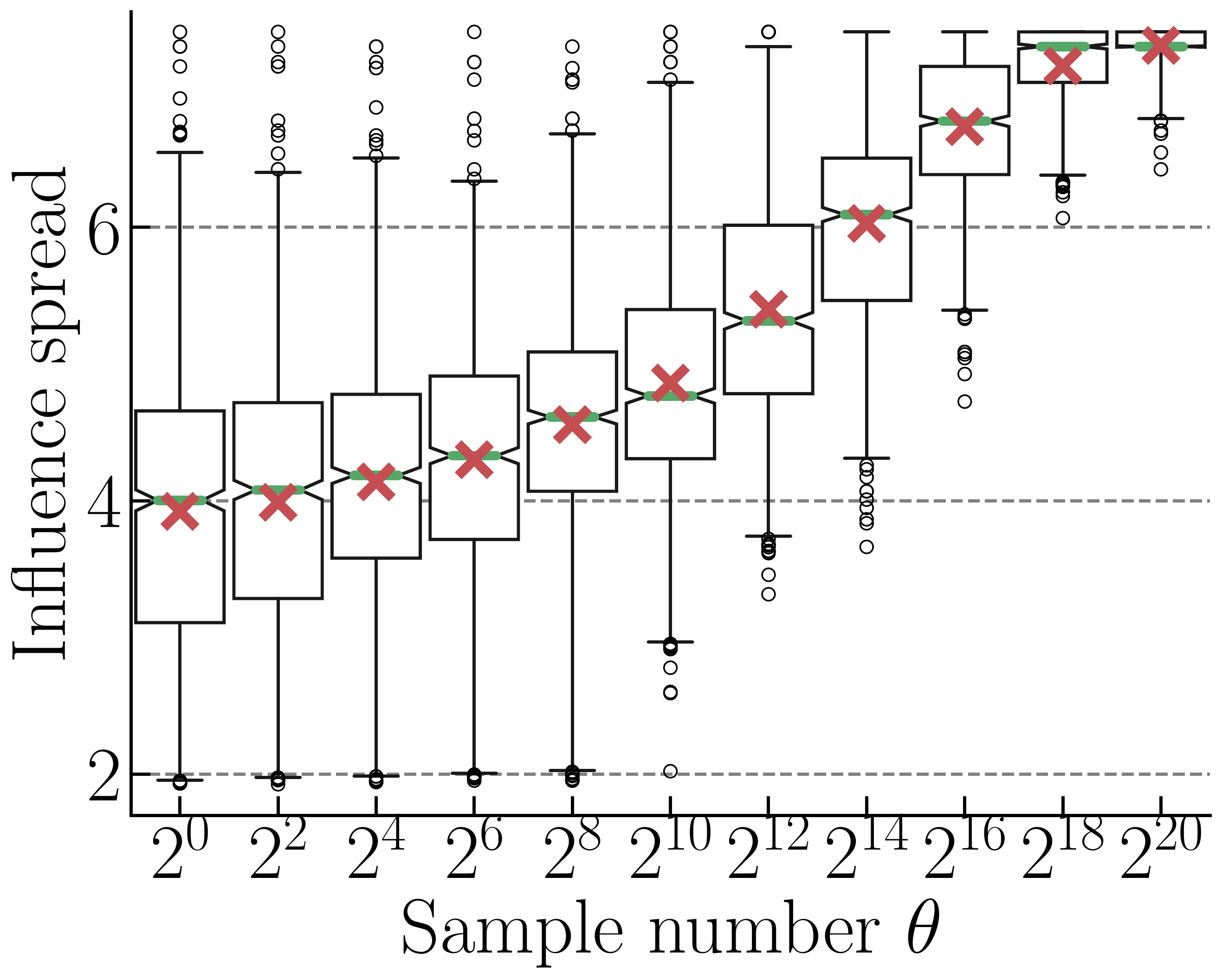}}
\caption{Influence distribution in notched box plot for \ris on \grqc ($k=1$).}
\label{fig:infdist:nbp-grqc}
\end{figure}

%% file: fig/mean_vs_std.tex
\begin{figure}[tbp]
\centering
\subfloat[Mean vs.~SD\label{fig:infdist:std}]{
	\includegraphics[width=0.5\hsize]{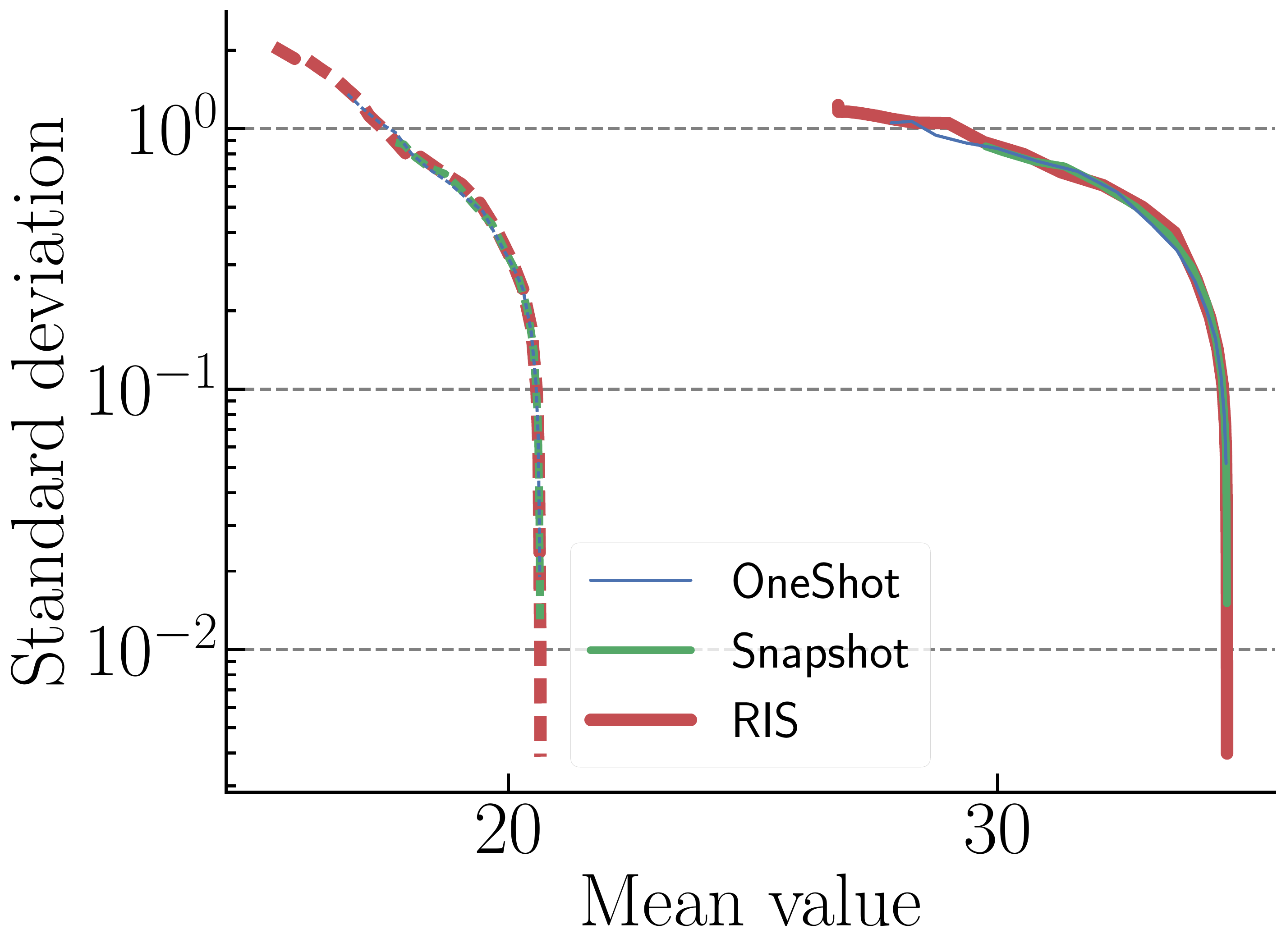}
}
\subfloat[Mean vs.~1st percentile\label{fig:infdist:skewness}]{
	\includegraphics[width=0.5\hsize]{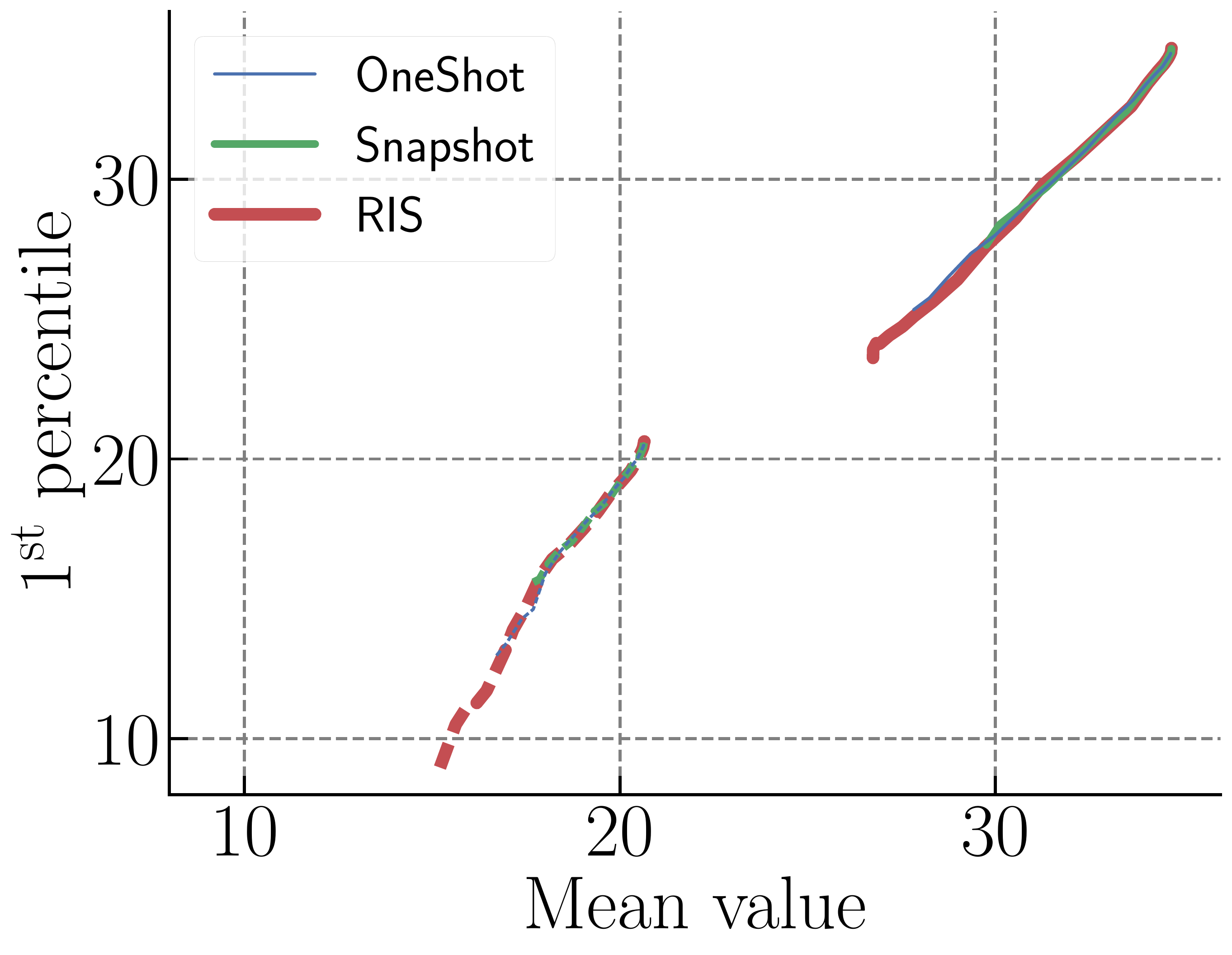}
}
\caption{Relation between mean value and other statistics of influence distributions, where
solid lines are for \phy (\OWC, $k=4$) and
dashed lines are for \phy (\UC{0.1}, $k=16$).
}
\end{figure}

%% file: fig/cr-ss-os.tex
\begin{figure}[tbp]
\subfloat[\phy (\UC{0.01})\label{fig:infdist:cr-ss-os:u001}]{
	\includegraphics[width=0.5\hsize]{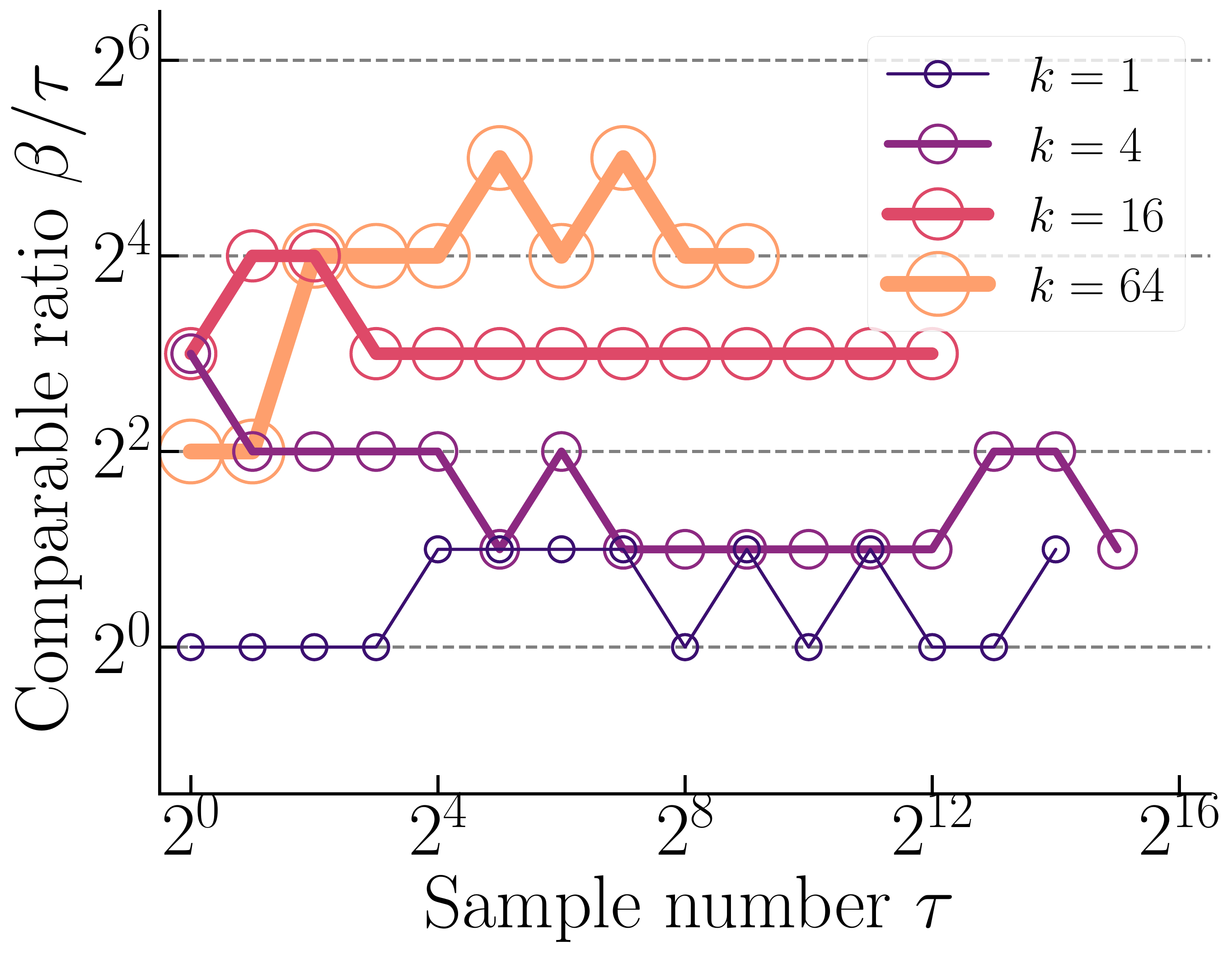}
}
\subfloat[\phy (\IWC)\label{fig:infdist:cr-ss-os:iwc1}]{
	\includegraphics[width=0.5\hsize]{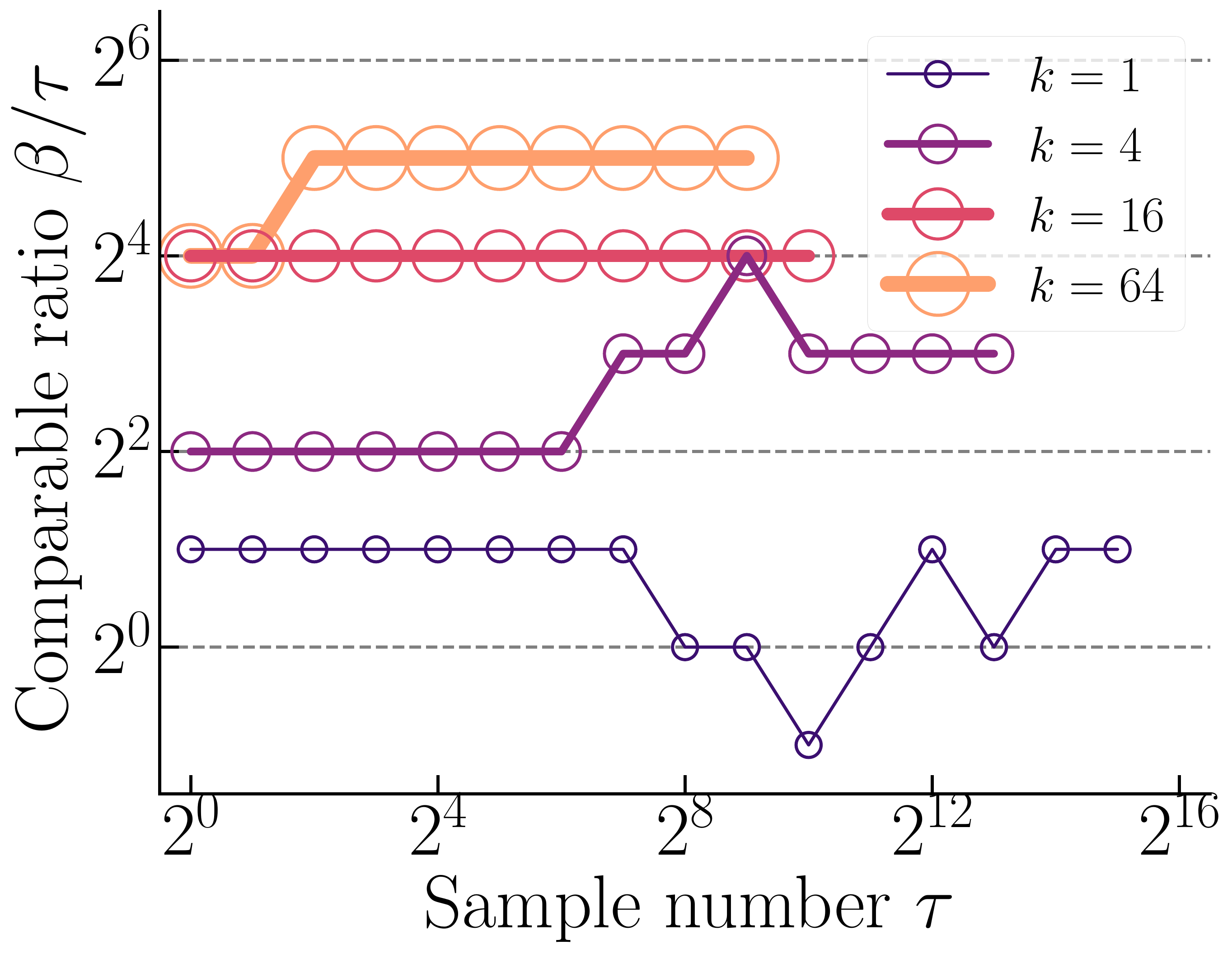}
}
\caption{Comparable number ratio of \oneshot to \snapshot.}
\label{fig:infdist:cr-ss-os}
\end{figure}

%% file: tab/median-cr-ss-os.tex
\begin{table}[tbp]
    \centering
    \caption{Median of comparable number ratio of \oneshot to \snapshot.
    }
    \label{tab:infdist:median-cr-ss-os}
\rowcolors{2}{gray!20}{white}
{\fontsize{7.5}{7.5}\selectfont
\setlength{\tabcolsep}{1pt}
\begin{tabular}{lr|rrrr}
\toprule
\textbf{network} & $k$ & \UC{0.1} & \UC{0.01} & \IWC & \OWC \\
\midrule
\karate &   1 &   2 &  1 &   2 &   2 \\
\karate &   4 &   8 &  6 &   8 &   8 \\
\karate &  16 &  16 &  8 &   8 &  12 \\
   \phy &   1 &   2 &  1 &   2 &   2 \\
   \phy &   4 &   4 &  3 &   6 &   4 \\
   \phy &  16 &   8 &  8 &  16 &  16 \\
$\star$
   \phy &  64 &  32 & 16 &  32 &  32 \\

  \grqc &   1 &   1 &  2 &   2 &   1 \\
  \grqc &   4 &   1 &  4 &   8 &   4 \\
  $\star$
  \grqc &  64 &   1 & 64 &  32 &  96 \\
  \wiki &   1 &  -- &  2 &   2 &   1 \\
  \wiki &   4 &  -- &  4 &   8 &   2 \\
   \BAs &   1 &   2 &  2 &   1 &   2 \\
   \BAs &   4 &   4 &  4 &   4 &   4 \\
   \BAs &  16 &   8 &  8 &   8 &   8 \\
   \BAd &   1 &   2 &  2 &   2 &   4 \\
   \BAd &   4 &   2 &  4 &   8 &  16 \\
   \BAd &  16 &  -- &  8 &  24 &  -- \\
\bottomrule
\end{tabular}
}
\end{table}

%% file: fig/cr-ss-ris.tex
\begin{figure}[tbp]
\centering
\subfloat[\phy (\UC{0.01})\label{fig:infdist:cr-ss-ris:u001}]{
	\includegraphics[width=0.5\hsize]{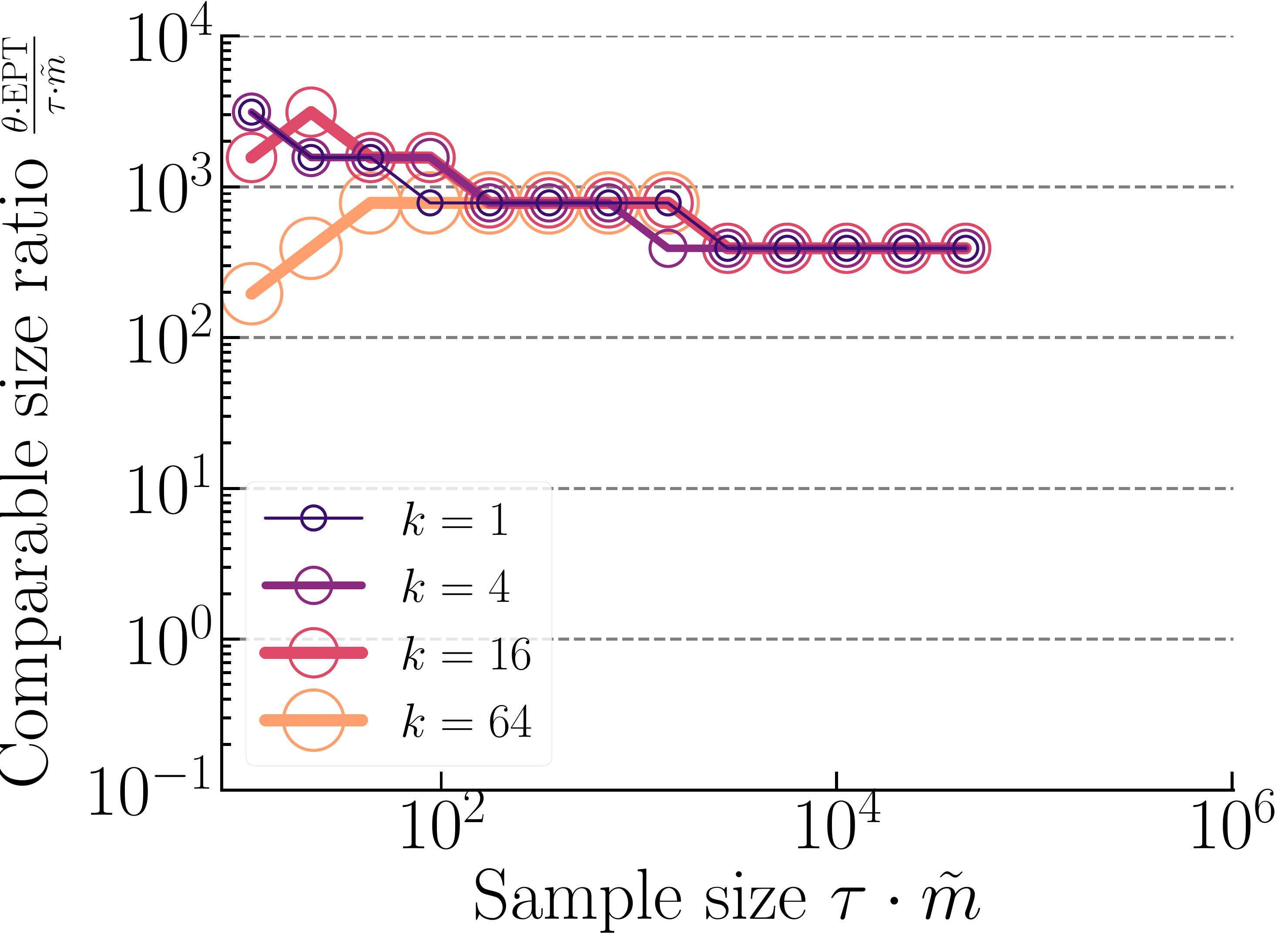}
}
\subfloat[\phy (\IWC)\label{fig:infdist:cr-ss-ris:iwc1}]{
	\includegraphics[width=0.5\hsize]{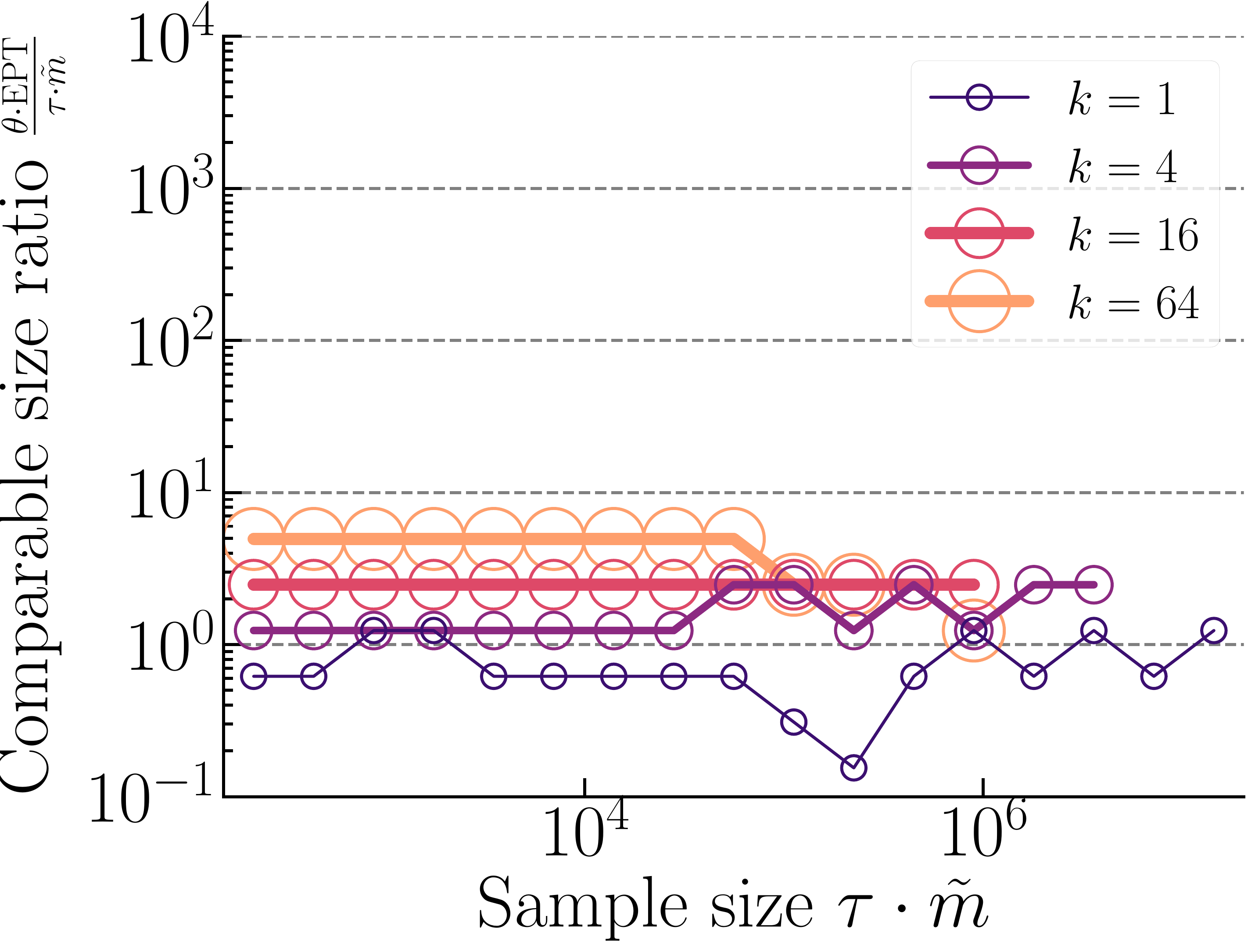}
}
\caption{Comparable size ratio of \ris to \snapshot.}
\label{fig:infdist:cr-ss-ris}
\end{figure}

%% file: tab/median-cr-ss-ris.tex
\begin{table}[tbp]
    \centering
    \caption{
    Median of comparable ratio of \ris to \snapshot.
    \hl{Number ratios} $>$ 4,096 and \hl{size ratios} $<$ 0.1 are highlighted.
    }
    \label{tab:infdist:median-cr-ss-ris}
\rowcolors{2}{gray!20}{white}
{\fontsize{7.5}{7.5}\selectfont
\setlength{\tabcolsep}{1pt}
\renewcommand{\arraystretch}{0.9}
\begin{tabular}{lr|rrrr|rrrr}
\toprule
& & \multicolumn{4}{c|}{number ratio $\theta / \tau$} & \multicolumn{4}{c}{size ratio $ (\theta \EPT) / (\tau \tilde{m}) $} \\
\textbf{network} & $k$ & \UC{0.1} & \UC{0.01} & \IWC & \OWC & \UC{0.1} & \UC{0.01} & \IWC & \OWC \\
\midrule
      \karate &     1 &    32 &         384 &          16 &          16 &    4 &   258 &      2 &      2 \\
      \karate &     4 &    32 &         512 &          16 &          32 &    4 &   344 &      2 &      3 \\
      \karate &    16 &   128 &       1,024 &          64 &          64 &   16 &   689 &      7 &      7 \\
         \phy &     1 &   256 &  \hl{8,192} &          32 &          64 &    4 &   782 &   0.62 &      1 \\
         \phy &     4 &   256 &  \hl{8,192} &          64 &          64 &    4 &   782 &      1 &      1 \\
         \phy &    16 &   256 &  \hl{8,192} &         128 &         128 &    4 &   782 &      2 &      2 \\
 $\star$ \phy &    64 &   512 &  \hl{8,192} &         256 &         256 &    8 &   782 &      5 &      5 \\
        \grqc &     1 &    32 &  \hl{8,192} &         256 &         768 & 0.13 &    30 &   0.19 &   0.57 \\
        \grqc &     4 &   128 & \hl{12,288} &         512 &       1,024 & 0.53 &    45 &   0.34 &   0.76 \\
$\star$ \grqc &    64 & 2,048 & \hl{65,536} &       1,024 &       2,048 &    9 &   241 &   0.76 &      2 \\
$\star$ \grqc & 1,024 &\hl{16,384} & \hl{98,304} &       4,096 &  \hl{8,192} &   68 &   362 &      3 &      6 \\
        \wiki &     1 &    -- &       2,048 &       1,024 &        512  &   -- &     2 &   0.75 &   0.31 \\
        \wiki &     4 &    -- &       2,048 &       1,024 &         768 &   -- &     2 &   0.75 &   0.46 \\
$\star$ \wiki &    64 &\hl{16,384} &  \hl{8,192} &       4,096 &       2,048 &  607 &    10 &      3 &      1 \\
$\star$ \wiki & 1,024 &\hl{49,152} & \hl{32,768} & \hl{32,768} & \hl{32,768} &1,820 &    40 &     24 &     20 \\
$\star$ \youtube &      1 &  -- &           128 &           64 &   \hl{65,536} &  -- &  0.13 &  \hl{0.00033} &        0.34 \\
$\star$ \youtube &      4 &  -- &           512 &        1,536 &   \hl{65,536} &  -- &  0.53 &   \hl{0.0080} &        0.34 \\
$\star$ \youtube &     16 &  -- &   \hl{32,768} &        2,048 &   \hl{65,536} &  -- &    34 &    \hl{0.011} &        0.34 \\
$\star$ \youtube &     64 &  -- &   \hl{65,536} &        4,096 &   \hl{65,536} &  -- &    68 &    \hl{0.021} &        0.34 \\
$\star$ \youtube &  1,024 &  -- &  \hl{393,216} &  \hl{32,768} &  \hl{524,288} &  -- &   409 &          0.17 &           3 \\
  $\star$ \pokec &      1 &  -- &   \hl{32,768} &        1,536 &         2,048 &  -- &  0.16 &    \hl{0.016} &  \hl{0.021} \\
  $\star$ \pokec &      4 &  -- &   \hl{24,576} &        1,536 &         1,536 &  -- &  0.12 &    \hl{0.016} &  \hl{0.016} \\
  $\star$ \pokec &     16 &  -- &   \hl{32,768} &        3,072 &         4,096 &  -- &  0.16 &    \hl{0.033} &  \hl{0.042} \\
  $\star$ \pokec &     64 &  -- &   \hl{81,920} &        4,096 &    \hl{8,192} &  -- &  0.41 &    \hl{0.043} &  \hl{0.085} \\
  $\star$ \pokec &  1,024 &  -- &  \hl{262,144} &  \hl{16,384} &   \hl{49,152} &  -- &     1 &          0.17 &        0.51 \\
         \BAs &     1 & 1,536 & \hl{32,768} &         512 &       1,024 &   17 & 3,314 &      2 &      4 \\
         \BAs &     4 & 2,048 & \hl{32,768} &         512 &       2,048 &   23 & 3,314 &      2 &      8 \\
         \BAs &    16 & 4,096 & \hl{65,536} &       1,024 &      2,048  &   46 & 6,627 &      4 &      8 \\
         \BAd &     1 &     4 &       2,048 &          16 &          24 & 0.54 &    21 &   0.24 &   0.36 \\
         \BAd &     4 &     8 &       2,048 &          32 &          48 &    1 &    21 &   0.48 &   0.72 \\
         \BAd &    16 &   512 &       4,096 &         128 &         256 &   69 &    43 &      2 &      4 \\
\bottomrule
\end{tabular}
}
\end{table}

%% file: tab/results-cost.tex
\begin{table*}[tbp]
    
\centering
\caption{Traversal cost at $k=1$ and sample number $1$ for each algorithm.
}
\label{tab:cost}
\rowcolors{2}{gray!20}{white}
{\fontsize{7.5}{7.5}\selectfont
\setlength{\tabcolsep}{2pt}
\begin{tabular}{ll|rr|rr|rr|rr}
\toprule
&
& \multicolumn{2}{c|}{\UC{0.1}}
& \multicolumn{2}{c|}{\UC{0.01}}
& \multicolumn{2}{c|}{\IWC}
& \multicolumn{2}{c}{\OWC}
\\
\textbf{network} & \textbf{algorithm} &
vertex & edge &
vertex & edge &
vertex & edge &
vertex & edge \\
\midrule
\karate &   \oneshot &       66.6 &        375.3 &      35.7 &      168.8 &      126.2 &        560.6 &      126.2 &      858.9 \\
$n=34$ &  \snapshot &       66.6 &         37.5 &      35.7 &        1.7 &      126.2 &        119.2 &      126.2 &      126.2 \\
$m=156$ & \ris &        2.0 &         11.0 &       1.1 &        5.0 &        3.7 &         25.3 &        3.7 &       16.5 \\ \midrule
\phy &   \oneshot &      429.9 &      2,008.8 &     252.5 &    1,153.2 &    1,020.8 &      4,636.5 &      986.6 &    4,700.4 \\
$n=241$ &  \snapshot &      429.9 &        200.9 &     252.5 &       11.5 &    1,020.9 &        904.3 &      986.5 &      900.4 \\
$m=1{,}098$ & \ris &        1.8 &          8.3 &       1.1 &        4.8 &        4.2 &         19.3 &        4.1 &       18.4 \\ \midrule
\grqc & \oneshot &   63,248.7 &  1,247,121.3 &   5,595.5 &   35,844.5 &   20,373.7 &    129,789.9 &   20,373.2 &  220,480.0 \\
$n=5{,}242$ & \snapshot &   63,210.5 &    124,630.7 &   5,595.4 &      358.4 &   20,374.3 &     19,625.0 &   20,377.5 &   20,377.4 \\
$m=28{,}968$ & \ris &       12.1 &        237.9 &       1.1 &        6.8 &        3.9 &         42.1 &        3.9 &       24.8 \\ \midrule
\wiki &   \oneshot &         -- &           -- &   8,959.0 &  184,956.1 &   12,449.6 &    233,880.9 &   26,242.1 &  924,225.6 \\
$n=7{,}114$ &  \snapshot &         -- &           -- &   8,959.0 &    1,849.6 &   12,449.7 &      5,365.7 &   26,243.6 &   19,385.0 \\
$m=103{,}689$ &       \ris &         -- &           -- &       1.3 &       26.0 &        1.8 &         46.3 &        3.7 &       46.8 \\ \midrule
$\star$ \youtube &  \snapshot &         -- &           -- &  70,630,278.9 &  77,780,873.5 &   6,713,554.8 &   6,133,043.3 &   6,712,576.5 &   6,712,316.1 \\
$n=$1.1M, $m=$6.0M &       \ris &         -- &           -- &          62.2 &       6,851.5 &           5.9 &       2,360.7 &           5.9 &          35.0 \\
\midrule
$\star$ \pokec &  \snapshot &         -- &           -- &   2,481,201.0 &     855,397.3 &  26,287,270.5 &  25,490,537.8 &  24,272,264.1 &  23,489,755.3 \\
$n=$1.6M, $m=$31M &       \ris &         -- &           -- &           1.5 &          52.2 &          16.1 &         890.8 &          14.9 &         325.1 \\
\midrule
\BAs & \oneshot &    1,131.8 &      1,318.0 &   1,010.2 &    1,026.8 &    2,276.0 &      1,844.1 &    2,233.1 &    3,987.3 \\
$n=1{,}000$ &  \snapshot &    1,131.8 &        131.8 &   1,010.3 &       10.3 &    2,276.2 &      1,276.2 &    2,233.4 &    1,233.4 \\
$m=999$ & \ris &        1.1 &          1.3 &       1.0 &        1.0 &        2.3 &          4.1 &        2.2 &        1.8 \\ \midrule
\BAd & \oneshot &  146,555.6 &  2,054,009.7 &   1,134.1 &   13,416.0 &   14,992.8 &    162,203.9 &   15,078.3 &  263,505.4 \\
$n=1{,}000$ & \snapshot &  146,577.9 &    205,464.5 &   1,134.0 &      134.1 &   15,014.4 &     14,924.5 &   15,055.5 &   15,053.0 \\
$m=10{,}879$ & \ris &      146.5 &      2,055.1 &       1.1 &       13.4 &       15.0 &        263.4 &       15.1 &      163.2 \\
\bottomrule
\end{tabular}
}

\end{table*}

%% file: discussion.tex
\clearpage

\section{Discussions on Traversal Cost}
\label{sec:disucsion}

\newcommand{\cros}{\mathrm{cr}_1}
\newcommand{\crris}{\mathrm{cr}_2}

We finally discuss the graph traversal cost
\emph{conditioning} that
the sample number is carefully specified so that
\oneshot, \snapshot, and \ris yield seed set distributions of identical accuracy.
Let $\cros$ and $\crris$ be
the comparable number ratios of \oneshot and \ris to \snapshot, respectively.
Setting
$\beta = \cros \cdot \gamma$,
$\tau = \gamma$, and
$\theta = \crris \cdot \gamma$
for any $\gamma$ ensures that
the three algorithms yield the-same-mean influence distributions.
Referring to Tables~\ref{tab:infdist:median-cr-ss-os}, \ref{tab:infdist:median-cr-ss-ris}, and \ref{tab:cost},
we can estimate the traversal cost when this is the case,
which are reported in Table~\ref{tab:discuss:grqc}.

\oneshot is almost always the least time-efficient.
Comparing between \snapshot and \ris,
we observe that \ris significantly surpasses \snapshot on
the two largest networks, \youtube and \pokec.
On the other hand,
\snapshot runs more than twice as fast as \ris on
\grqc (\UC{0.01}) and \BAs (\UC{0.1}, \UC{0.01}).
On such instances,
edge probabilities are too low, or the underlying graph is quite sparse and small,
resulting in large comparable ratios
(see Table~\ref{tab:infdist:median-cr-ss-ris}).

Summarizing the above,
we conclude that
\emph{as long as} we use naive implementations,
\textbf{1.}~either \snapshot or \ris is preferable;
\oneshot can be promising only if the size of available memory is limited, and
\textbf{2.}~\ris is more time-efficient than \snapshot for large complex networks;
conversely,
\snapshot is preferable for small, low-influence-probability networks.

\paragraph{Remarks}
Our discussion would need to be adapted for existing efficient implementations, and
we should note the following two facts.
First, there is a variety of boosted implementations designed for \oneshot and \snapshot.
Some of them \emph{significantly reduce the traversal cost without
sacrificing approximation guarantee}, e.g.,
SKIM proposed by Cohen, Delling, Pajor, and Werneck~\cite{cohen2014sketch}
is \snapshot-type
and guaranteed to run in near-linear time $ \bigO( \tau(n+m) + m \epsilon^{-2} \log^2 n ) $
(for some $\epsilon$ independent of sample number).
Second,
none of the existing \oneshot- and \snapshot-type algorithms
adopted bounds on the sample number in Sections~\ref{sec:review:one-shot} and~\ref{sec:review:snapshot}.
This situation motivates applying
(tight) bounds on the sample number to \oneshot and \snapshot.

\input{tab/discuss-grqc}

%% file: tab/discuss-grqc.tex
\begin{table}[tbp]
\rowcolors{2}{gray!20}{white}
\centering
\caption{Traversal cost at $k=1$ conditioning that \oneshot, \snapshot, and \ris are of identical accuracy.
}
\label{tab:discuss:grqc}
{\fontsize{7.5}{7.5}\selectfont
\setlength{\tabcolsep}{1pt}
\begin{tabular}{cc|rrrr}
\toprule
\textbf{network} & \textbf{algorithm} & \UC{0.1} & \UC{0.01} & \IWC & \OWC \\
\midrule
\grqc & \oneshot &  1,310,453$\gamma$ &       82,882$\gamma$ &     300,318$\gamma$ &     240,829$\gamma$ \\
\grqc &  \snapshot &    187,970$\gamma$ &        \mini{5,954$\gamma$} &      39,997$\gamma$ &      40,744$\gamma$ \\
\grqc &       \ris &      \mini{8,000$\gamma$} &       64,757$\gamma$ &      \mini{11,762$\gamma$} &      \mini{22,000$\gamma$} \\
\midrule
\wiki & \oneshot &                 -- &      387,830$\gamma$ &     492,661$\gamma$ &     950,468$\gamma$ \\
\wiki & \snapshot &                 -- &       \mini{10,809$\gamma$} &      \mini{17,815$\gamma$} &      45,629$\gamma$ \\
\wiki &      \ris &                 -- &       55,830$\gamma$ &      49,202$\gamma$ &      \mini{25,859$\gamma$} \\
\midrule
$\star$ \youtube & \snapshot &                 -- &  148,411,152$\gamma$ &  12,846,598$\gamma$ &  13,424,893$\gamma$ \\
$\star$ \youtube &      \ris &                 -- &      \mini{884,953$\gamma$} &     \mini{151,464$\gamma$} &   \mini{2,679,453$\gamma$} \\
\midrule
$\star$ \pokec &  \snapshot &                 -- &    3,336,598$\gamma$ &  51,777,808$\gamma$ &  47,762,019$\gamma$ \\
$\star$ \pokec &       \ris &                 -- &    \mini{1,760,893$\gamma$} &   \mini{1,393,038$\gamma$} &     \mini{696,292$\gamma$} \\
\midrule
\BAs &   \oneshot &      4,899$\gamma$ &        4,074$\gamma$ &       4,120$\gamma$ &      12,441$\gamma$ \\
\BAs &  \snapshot &      \mini{1,264$\gamma$} &        \mini{1,021$\gamma$} &       3,552$\gamma$ &       \mini{3,466$\gamma$} \\
\BAs &       \ris &      3,762$\gamma$ &       66,751$\gamma$ &       \mini{3,282$\gamma$} &       4,154$\gamma$ \\
\midrule
\BAd &   \oneshot &  4,400,487$\gamma$ &       29,100$\gamma$ &     354,374$\gamma$ &   1,114,474$\gamma$ \\
\BAd &  \snapshot &    351,923$\gamma$ &        \mini{1,268$\gamma$} &      29,902$\gamma$ &      30,176$\gamma$ \\
\BAd &       \ris &      \mini{8,806$\gamma$} &       29,798$\gamma$ &       \mini{4,454$\gamma$} &       \mini{4,278$\gamma$} \\
\bottomrule
\end{tabular}
}
\end{table}

%% file: conclusion.tex
\section{Concluding Remarks}
\label{sec:conclusion}

In this paper, we established an experimental study on three algorithmic approaches for influence maximization.
Possible directions for developing of new algorithms are listed below.

\begin{itemize}[leftmargin=*]
    \item
    \textbf{Sample number selection for \oneshot and \snapshot}:
    If the sample number is specified so that the three algorithms are of identical accuracy,
    then \oneshot can be the most space-saving and
    \snapshot can have less expensive traversal costs than \ris for small graphs.
    Can we \emph{efficiently} calculate a tight bound on the sample number for \oneshot and \snapshot (in Sections~\ref{sec:review:one-shot} and~\ref{sec:review:snapshot}), or apply
    \ris's sample number determination to \oneshot and \snapshot?

    \item
    \textbf{Space reduction for \snapshot and \ris}:
    Both \snapshot and \ris may consume much memory despite their time efficiency.
    Can we cut down the memory usage of \snapshot and \ris,
    e.g., by compressing reverse-reachable sets?
\end{itemize}

We also give two aspects of experimental evaluation.
Only carrying out either of them would not be enough.
    One is to test the trade-off between scalability and actual influence with \emph{varying sample numbers} as in this paper.
    One can identify the fastest algorithm
    on the assumption that the sample number is ideally chosen.
    Note that \ris-type algorithms would show
    nearly identical trade-off~\cite[Sect.~3.1]{lu2017refutations}.
    The other is to test the scalability of algorithms for a
    \emph{given precision requirement}
    (e.g., a $(1-1/\rme-\epsilon)$-approximation with probability $1-\delta$).
    One can identify the most efficient algorithm taking into account the sample number selection.
    Currently, existing \oneshot- and \snapshot-type algorithms would not be applicable for this test
    due to the lack of a mechanism for sample number selection.

\section*{Acknowledgments}
The author would like to thank anonymous reviewers for their constructive comments and suggestions and
was supported by JST ERATO Grant Number JPMJER1201, Japan.

\section*{Appendix}

\textbf{Traversal cost for \oneshot at $k=1$.}
Since a vertex is scanned if it is activated, the vertex traversal cost is equal to
$\sum_{v} \Inf_\calG(v)$.
For each scanned vertex,
we touch its out-going edges;
we then have that the edge traversal cost is

$ \sum_{v} \E_G[\sum_{w: v \text{ can reach } w} \odeg_G(w)]
= \sum_{w} \odeg(w) \cdot \E_G[r_{G^\top}(w)] \\
= \sum_{w} \odeg(w) \cdot \Inf_{\calG^\top}(w)
\leq m \cdot \max_{v} \Inf_{\calG^\top}(v).$

\noindent
\textbf{Sample size of \snapshot and \ris.}
By definition, it follows that
$\sum_{e} p(e) = \E_{G}[|E(G)|]$.
Observing that $r_G(v) \leq 1+|E(G)|$ for all $v$,
we have that
$
\frac{1}{n} \sum_{v} \Inf(v) =
\frac{1}{n} \sum_{v} \E_G[r_G(v)] \leq
\frac{1}{n} \sum_{v} \E_G[1+|E(G)|] =
1 + \sum_{e} p(e).
$